\documentclass[sigconf,nonacm]{acmart}
\usepackage{popets}

\usepackage{bm}
\usepackage{array,multirow,tabularx}
\usepackage{makecell}
\usepackage{algorithm}
\usepackage{algpseudocode}
\usepackage{float}

\algnewcommand\algorithmicforeach{\textbf{for each}}
\algdef{S}[FOR]{ForEach}[1]{\algorithmicforeach\ #1\ \algorithmicdo}
\algrenewcommand\textproc[1]{\textit{#1}}

\setcopyright{none}
\copyrightyear{2027}
\acmYear{2027}
\acmVolume{0}
\acmNumber{0}
\acmDOI{}
\acmConference[Preprint]{Preprint}{}{}
\acmISBN{}
\keywords{Prompt privacy, large language models, agent systems, OCR-mediated multimodal inputs, privacy-preserving inference, sensitive entity detection, prompt sanitization, utility preservation}

\begin{document}

\title{BodhiPromptShield: Pre-Inference Prompt Mediation for Surface-Form Privacy Propagation in LLM Agent Pipelines}

\author{Bo Ma}
\affiliation{%
  \institution{Auckland University of Technology}
  \city{Auckland}
  \country{New Zealand}}
\email{rcn4743@aut.ac.nz}

\author{Jinsong Wu}
\affiliation{%
  \institution{Guilin University of Electronic Technology}
  \city{Guilin}
  \country{China}}
\email{wujs@ieee.org}

\author{Weiqi Yan}
\affiliation{%
  \institution{Auckland University of Technology}
  \city{Auckland}
  \country{New Zealand}}
\email{weiqi.yan@aut.ac.nz}

\begin{abstract}
In LLM agent pipelines, prompt privacy risk propagates beyond a single model call: raw user content enters retrieval queries, memory writes, tool arguments, OCR-derived text, and logs, and every downstream copy inherits what the first write contained. Existing de-identification pipelines protect document boundaries but not this cross-stage surface. We present BodhiPromptShield, a policy-aware mediation layer that detects sensitive spans before they propagate, replaces each with a typed placeholder, a semantic abstraction, or a secure symbolic token under a configured policy, and defers restoration to authorized execution boundaries. We evaluate it under one protocol against Presidio, Casper-style sanitization, an LLM sanitizer, and transformer and learned detectors, on 300 AI4Privacy documents, 493 PrivacyLens trajectories, 200 PrivacyLens tasks scored by that benchmark's own judge, and AgentDojo tasks under injection. Three findings result. Identifier propagation is controllable: residual exposure falls to 7.4\% on AI4Privacy and 1.8\% on PrivacyLens, and exact identifiers in an agent's final action fall from 13.7\% to 2.1--3.1\%. Restoration timing governs what every stage upstream of the authorized boundary sees: deferring it leaves 1.6\% of protected values readable in the released context against 51.0\%, and 2.7\% against 4.8\% in what the agent emits, for 0.11 helpfulness points. Measuring factual disclosure is harder: a word-overlap metric and an LLM judge both report that mediation leaves facts intact, and both disagree with blind human annotation ($\kappa=0.25$ and $0.09$). The human labels reverse that: inferability falls from 100\% to 24--53\% under mediation, so semantic-leakage measures need human validation before they are trusted. These are systems results on English text with open-weight models, not formal guarantees.
\end{abstract}

\maketitle

\begin{center}\small\emph{Preprint. This paper is under submission; it has not been peer reviewed and no venue is asserted.}\end{center}

\section{Introduction}
Large language models (LLMs) now drive many interactive assistants and tool-using agent pipelines~\cite{yao2022react,schick2023toolformer}. In these systems, users frequently submit prompts containing sensitive information, such as identifiers, financial data, medical content, and private text extracted from OCR-mediated images.

This creates a practical exposure problem at inference time. Raw prompts can propagate through logging, retrieval, tool calls, memory stores, and third-party components. Consequently, unmediated prompts may leak sensitive content even when training-time privacy is not the immediate issue~\cite{carlini2021extracting,mireshghallah2023privacy}. Agent orchestration further enlarges this surface because prompt content is repeatedly transformed and relayed across components, and prompt-mediated attacks can exploit these paths~\cite{perez2022ignore,greshake2023not}.

Most prior privacy-preserving NLP work emphasizes training-time protection, including differential privacy, representation anonymization, and post hoc de-identification~\cite{dwork2014algorithmic,chaudhuri2011differentially}. These methods remain important, but they do not directly solve interface-layer exposure in deployed LLM systems. Conventional enterprise redaction middleware likewise focuses on ingress scrubbing or stored-text de-identification. That framing is too weak for modern agents: a system can mask a span once and still leak it if partially protected content is copied into retrieval queries, cached in memory, restored too early, or forwarded as tool arguments.

A concrete deployment example is a reimbursement assistant that receives an invoice image, extracts text with OCR, retrieves a company policy, stores a running plan in memory, and finally calls an expense tool. If the invoice ID, supplier address, or account reference enters the pipeline in raw form, those values can be duplicated across retrieval prompts, memory entries, and tool logs before the final execution step. A document-boundary de-identification score does not measure this failure mode; what matters is whether protected content stays suppressed across agent states until an authorized boundary actually needs the exact value. Assistants of this shape are in production now, and the operator answerable for what crosses their boundary needs a measurement at each hop.

Enterprise tenant isolation, scoped retrieval indexes, and role-based tool permissions reduce cross-user exposure, but they do not by themselves answer this within-agent question. A correctly authorized agent can still copy the same user's sensitive span into an internal query, memory trace, tool argument, or audit log before any later redaction step runs. This residual threat is the focus of the paper: propagation within a partially trusted pipeline, not provider-side training retention or cross-tenant access-control failure.

This gap motivates a sharper research question than ordinary span masking: can privacy mediation suppress explicit surface-form propagation across agent states while retaining enough semantic structure for reliable downstream reasoning and execution? Existing de-identification metrics can say whether a span was transformed at a document boundary; they cannot say whether privacy survives agent-state transitions. The contribution is deployment-oriented rather than guarantee-oriented---we claim no new cryptographic or differential-privacy mechanism---and the setting is the one organizations are deploying now: assistants that read a user's document, search an internal index, keep a running memory, and call a tool that writes to a real system. In such a stack the difference between masking once and controlling propagation is the difference between a redacted prompt and a sensitive identifier sitting in a retrieval log, a vector store, and a third-party tool's audit trail. The measurements in this paper are about what an operator of such a system can and cannot expect a mediation layer to prevent.

The main contributions of this paper are summarized as follows:
\begin{itemize}
    \item We reformulate prompt privacy in LLM agent pipelines as a propagation-control problem rather than a single-document masking problem, and map each threat in the model to a quantity that can be measured at an agent boundary.
    \item We treat restoration timing as an explicit security variable: deferring restoration to the authorized execution boundary leaves 1.6\% of protected values readable in the released context against 51.0\% when they return during planning, and cuts the identifiers the agent itself writes before that boundary by a paired $2.1$ points [$0.8$, $3.6$], for 0.11 points of task helpfulness.
    \item We compare the mechanism head-to-head under one protocol against installed Presidio, Casper-style sanitization, an LLM sanitizer, and transformer and learned detectors, on real agent data with confidence intervals on every reported quantity.
    \item We separate surface-form leakage from factual leakage and validate the measures for the second against blind human annotation. Both automated proxies fail that check, and the human labels show mediation cutting factual inferability from 100\% to 24--53\%---a result neither proxy reports.
\end{itemize}

The rest of this paper is organized as follows. Section~\ref{sec:problem} presents the problem definition and threat model. Section~\ref{sec:related} reviews related work. Section~\ref{sec:framework} presents the proposed prompt mediation framework. Sections~\ref{sec:setup} and~\ref{sec:results} describe the controlled evaluation protocol and discuss the results. Section~\ref{sec:limitations} outlines limitations, and Section~\ref{sec:conclusion} concludes the paper.

\section{Problem Definition and Threat Model}
\label{sec:problem}
\subsection{Problem Definition}
Let a raw user prompt be denoted by $x$. In the deployment setting considered here, $x$ enters an agent pipeline that can be abstracted as a directed boundary graph
\[
G=(V,\mathcal{E}),
\]
whose nodes represent prompt ingress, retrieval, memory, planning, tool execution, and logging boundaries. The prompt may contain a set of privacy-sensitive spans
\[
S(x)=\{s_1,s_2,\ldots,s_n\}.
\]
Each span $s_i$ may correspond to a privacy category such as personally identifiable information, financial information, health-related information, confidential organizational terms, or sensitive text extracted from visual input. After extraction, the working annotation is
\[
E(x)=\{(s_i,c_i,p_i)\}_{i=1}^{n},
\]
where $c_i$ is the privacy category and $p_i\in[0,1]$ is the detection confidence used by the routing policy.

The system applies a policy-instantiated mediation operator
\[
T_{\pi}: (x,E(x))\mapsto (\tilde{x},K),
\]
where $\pi$ denotes a deployment policy profile, $\tilde{x}$ is the sanitized prompt released to the downstream agent, and $K$ is an optional secure mapping table kept outside ordinary agent state. The problem is therefore not merely to rewrite text once at a document boundary, but to choose a mediation policy that keeps privacy-bearing surface forms from propagating across internal agent edges before an authorized execution step.

The mediated system should satisfy the following properties:
\begin{enumerate}
    \item privacy-sensitive content in $x$ is hidden, generalized, replaced, or symbolically mapped in $\tilde{x}$;
    \item explicit surface-form propagation along internal agent boundaries remains low before any authorized restoration event;
    \item the semantic intent of the original prompt is preserved as much as possible for downstream reasoning and execution.
\end{enumerate}
Given a downstream model $f(\cdot)$, the revised objective is not conventional text classification, but privacy-aware inference:
\[
y=f(\tilde{x}),
\]
where $\tilde{x}$ is generated from $x$ through pre-inference prompt mediation before the agent can branch into retrieval, memory, or tool-use stages.

In some scenarios, an authorized post-processing or secure recovery mechanism may be used. We denote this by
\[
\mathcal{R}_{\rho}(y,K)\mapsto y',
\]
where $\rho$ is the restoration policy and $K$ enables partial recovery of protected entities only at approved execution boundaries. The central problem formulation of the paper is therefore: choose $(\pi,\rho)$ so that direct prompt exposure, explicit cross-boundary surface-form propagation, and downstream utility loss are jointly controlled in the same agent workflow. This paper studies one half of that pair. It fixes $\pi$ per evaluated condition and treats the restoration policy $\rho$ as the variable, measuring what moving it costs and buys; jointly optimizing the two is the open problem the formulation names rather than a result reported here.

\subsection{Threat Model}
We consider a partially trusted deployment environment for LLM-based agents and separate two surfaces that are often conflated. The first is the deployed agent pipeline: retrieval queries, memory writes, tool arguments, OCR-derived text, and logs may expose raw user content to downstream observers. The second is the mediation system itself: symbolic tokens, restoration credentials, and the mapping table can become targets if exact-value restoration is enabled.

Within that setting we distinguish four adversary classes. An \emph{honest-but-curious downstream observer} may inspect released agent traces if mediation is not applied before inference. An \emph{adaptive prompt adversary} may craft prompts to evade span detection or trigger unsafe restoration through paraphrases, Unicode/homoglyph substitutions, mixed-language mentions, or prompt-injection-style instructions~\cite{perez2022ignore,greshake2023not,zou2023universal}. A \emph{mapping-table adversary} may target the secure mapping table, restoration tokens, or audit/logging infrastructure. Finally, a \emph{context-inference adversary} may infer private attributes from non-redacted context even when direct identifiers are hidden.

The attacker is therefore assumed to observe one or more of those surfaces, the plaintext prompt itself, or the restoration metadata. The attacker aims either to recover protected spans directly or to infer private attributes from contextual evidence and downstream traces~\cite{staab2024beyond}. Under this view, privacy protection should occur before prompt forwarding, rather than only during offline training or after response generation.

We do not assume that the full downstream stack is cryptographically secure, and we do not claim robustness against arbitrary host compromise or training-time poisoning. Instead, this paper focuses on a practical deployment setting in which prompt-level privacy mediation reduces explicit exposure, makes the security assumptions visible, and remains compatible with existing LLM agent systems.

\begin{table*}[t]
\centering
\caption{Threat-to-evaluation mapping. The table separates adversary capability, protected surface, measured evidence, and the strength of the claim supported by the released artifacts.}
\label{tab:threat_map}
\scriptsize
\setlength{\tabcolsep}{3pt}
\begin{tabularx}{\textwidth}{>{\raggedright\arraybackslash}p{2.5cm}>{\raggedright\arraybackslash}X>{\raggedright\arraybackslash}X>{\raggedright\arraybackslash}X>{\centering\arraybackslash}p{1.55cm}}
\toprule
Threat class & Capability & Protected surface & Current evidence & Claim status \\
\midrule
Honest-but-curious downstream observer & Reads retrieval queries, memory writes, tool arguments, logs, or OCR-derived text after prompt ingress & Explicit sensitive surface forms released across agent boundaries & PER and SPE across matched retrieval, memory, and tool stages & Direct \\
Adaptive prompt adversary & Perturbs or reframes sensitive spans using homoglyphs, paraphrase, code switching, or restoration-trigger instructions & Span detector and restoration policy & Executed surface-form evasion probes plus restoration-trigger probes & Partial \\
Mapping-table adversary & Reads symbolic tokens, restoration metadata, partial table rows, or restoration credentials & Secure mapping table $K$, per-session token namespace, and authorized restoration boundary & Executed reference sanity check on unauthorized restoration, partial table read, and cross-session token linkage; full table compromise is out of scope & Partial/design \\
Context-inference adversary & Infers sensitive attributes from remaining non-identifier context & Latent attributes not directly represented as surface spans & Executed attribute-recovery attack on the released surface, read beside literal survival, plus blind human annotation & Partial \\
\bottomrule
\end{tabularx}
\end{table*}

Table~\ref{tab:threat_map} keeps the threat model aligned with the manuscript's actual empirical scope. Honest-but-curious boundary observation is directly evaluated. Adaptive evasion, unsafe restoration, and context inference are only partially assessed, and a mapping-table compromise is treated as a deployment failure for the affected sessions rather than as an attack the current prototype can withstand. This distinction is important for interpreting the results: the paper measures explicit surface-form exposure and propagation, not a general bound on semantic recovery.

\subsection{Design Objective}
The framework should balance three coupled quantities: direct residual exposure in the released prompt, explicit surface-form propagation across agent boundaries, and downstream utility. We write the deployment objective as a conceptual systems objective
\[
\min_{\pi,\rho,T_{\pi}}\;
\mathcal{L}_{\text{direct}}(x,\tilde{x})
+\gamma R_{\text{prop}}(x;\pi,\rho)
+\lambda\mathcal{L}_{\text{utility}}(f(x),f(\tilde{x})),
\]
where $\mathcal{L}_{\text{direct}}$ measures residual direct exposure after mediation and $\mathcal{L}_{\text{utility}}$ captures semantic or task degradation. The middle term is the deployment-level propagation risk over agent edges, $R_{\text{prop}}(\pi,\rho)=\sum_{e\in\mathcal{E}} w_e q_e^{\pi,\rho}$, where $q_e^{\pi,\rho}$ is the probability that privacy-bearing content traverses edge $e$ under sanitization policy $\pi$ and restoration policy $\rho$, and $w_e$ is that boundary's exposure weight. Mediation is intended to lower $q_e^{\pi,\rho}$ before the first downstream branching point, and late restoration to confine exact-value reintroduction to a small subset of authorized execution edges. The weights $\gamma$ and $\lambda$ are deployment hyperparameters rather than theorem constants. The current experiments do not estimate the full scalar $R_{\text{prop}}$, because that would require edge traversal probabilities and policy-dependent restoration probabilities in a live agent stack. Instead, they report PER for direct exposure, SPE for explicit surface-form survival at each released boundary, and AC/TSR/UPR for utility under matched downstream settings.

\subsection{Formal Privacy Scope}
\label{sec:formalscope}
The evaluated pipeline is deterministic and provides no end-to-end differential-privacy guarantee. Even a randomized span-replacement layer would not cover untouched context, downstream outputs, prompt-injection bypasses, semantic inference, or compromise of the secure mapping table. We therefore report PER, SPE, FRR, and attack success rates as empirical measurements with explicit scope, not as privacy, recovery, or inference bounds. A formal end-to-end guarantee would require a different mechanism and a proof covering the complete agent state and restoration path, which is outside this work.

\section{Related Work}
\label{sec:related}
\subsection{Privacy-Preserving NLP and De-identification}
Privacy-preserving NLP has historically emphasized protection of training corpora and released datasets. Representative directions include differential privacy in optimization and embedding learning~\cite{dwork2014algorithmic,chaudhuri2011differentially}, as well as rule-based and statistical de-identification pipelines for clinical text~\cite{neamatullah2008automated,dernoncourt2017deidentification,uzuner2007evaluating,meystre2010extracting}. Foundational de-identification frameworks such as k-anonymity and practical health-data de-identification guidance further shaped this literature~\cite{sweeney2002kanonymity,elemam2011systematic}. However, these strands primarily target stored records and model development artifacts, and do not directly address real-time prompt exposure in interactive agent systems.

\subsection{Large-Model Privacy and Prompt-Level Risks}
Large-model deployment introduces privacy risk channels beyond training-time memorization. Prompt content may be exposed through service logs, retrieval requests, memory traces, and third-party integrations~\cite{mireshghallah2023privacy,weidinger2021ethical}. Training-data extraction, membership-inference, and inferential privacy studies demonstrate model-side leakage potential~\cite{carlini2021extracting,shokri2017membership,staab2024beyond,lukas2023analyzing}, while prompt-injection and jailbreak studies show that inference-time instruction channels can be adversarially manipulated~\cite{perez2022ignore,greshake2023not,zou2023universal,weiss2023jailbroken}. These results jointly motivate pre-inference prompt mediation under realistic deployment assumptions.

\subsection{Closest Prompt-Sanitization Systems}
The closest prior systems make the novelty boundary important. Casper is an on-device browser extension that combines rules, a local NER component, and a local topic identifier; it replaces detected PII with dummy values and restores them through a local mapping after the cloud response~\cite{chong2024casper}. PrivacyRestore removes privacy spans at the client and uses learned restoration vectors with activation steering during server-side inference~\cite{zeng2025privacyrestore}. Two 2026 systems narrow the gap: PromptGraph scores spans over an attributed graph and restores placeholders after local consistency checks~\cite{gu2026promptgraph}, and PAAC has a cloud agent reason over typed placeholders that an on-device registry substitutes and reverses~\cite{yuan2026paac}. Both settle sanitize-then-restore as prior art, and both restore within a single exchange rather than treating the restoration point as a variable. Presidio is a practical de-identification SDK~\cite{microsoft_presidio}. Our claim is narrower and different in object: we measure whether the mediated surface survives explicit retrieval, memory, and tool boundaries, and we treat the restoration boundary and per-session mapping as agent-governance variables. The replacement operation itself is not new.

\begin{table}[t]
\centering
\caption{Comparison with the closest prompt-sanitization and de-identification systems. Placeholder substitution and restoration are prior art; the evaluated distinction is agent-boundary propagation and restoration timing. Casper-style and Presidio are reproduced and compared under this paper's protocol; PrivacyRestore requires white-box inference control and is discussed rather than reproduced.}
\label{tab:prior_comparison}
\scriptsize
\setlength{\tabcolsep}{2.5pt}
\begin{tabularx}{\columnwidth}{>{\raggedright\arraybackslash}p{1.7cm}>{\raggedright\arraybackslash}X>{\raggedright\arraybackslash}X>{\raggedright\arraybackslash}X}
\toprule
System & Protection / restoration & Agent-boundary propagation & Evidence emphasis \\
\midrule
Casper~\cite{chong2024casper} & On-device rules, NER, topic identification, dummy-value replacement and local response restoration & Not evaluated across retrieval, memory, and tool surfaces; mechanism reproduced here as \emph{Casper-style} and measured on those surfaces & Prompt sanitization and user-side deployment \\
\makecell[l]{Privacy\\Restore~\cite{zeng2025privacyrestore}} & Client-side span removal and server-side activation-based restoration vectors & Client--server inference setting; requires white-box activation control, so not reproducible behind a black-box agent API & Privacy removal/restoration and inference utility \\
Presidio~\cite{microsoft_presidio} & Recognizer/anonymizer middleware with typed or configurable replacements & Single-boundary de-identification; no late restoration governance in this paper's protocol & Practical entity recognition and anonymization \\
This work & Typed placeholder, abstraction, or symbolic mapping with authorized late restoration & Explicit retrieval/memory/tool surface survival, fact retention, and mapping-boundary tests & Deployment-oriented propagation measurements; no formal guarantee \\
\bottomrule
\end{tabularx}
\end{table}

Table~\ref{tab:prior_comparison} separates the two comparisons the paper can
and cannot make empirically. Casper's mechanism---rule and NER detection,
dummy-value substitution, and local restoration through a client-held
mapping---is reimplemented and run head-to-head under the single protocol of
Section~\ref{sec:setup}, on both propagation sources and in the downstream
task-native evaluation; it appears in every results table as
\emph{Casper-style}. We label it \emph{-style} because it reproduces the
published mechanism rather than the released browser extension, whose topic
identifier and interaction surface are not part of an agent pipeline.
PrivacyRestore is not reproduced, and the reason is architectural rather than
one of effort: it restores privacy content by steering hidden activations
\emph{inside} the serving model, which presupposes white-box control of
inference. The mediation layer studied here sits in front of a model it treats
as a black box behind an API---the deployment position that makes
agent-boundary propagation a problem in the first place---so the two designs
occupy different trust boundaries and cannot be placed on one protocol
without changing what either is. We discuss PrivacyRestore as prior
art on client-side removal with authorized recovery and make no empirical
claim against it.

A second line of work protects prompts by rewriting or minimizing them with a model rather than a recognizer. PAPILLON routes a local model in front of a stronger remote one so that only a privacy-scrubbed prompt crosses the trust boundary~\cite{siyan2024papillon}, Rescriber puts a small local model behind a user-facing minimization loop for chatbot inputs~\cite{zhou2025rescriber}, and Dou et al.\ rewrite online self-disclosures to lower re-identification risk while preserving what the author meant to say~\cite{dou2024reducing}. These systems share our premise that mediation belongs at the interface rather than in training, and they motivate the LLM-sanitizer baseline evaluated in Section~\ref{sec:results}. They differ in target: each protects a single request-response exchange, whereas the question here is what survives once that exchange is decomposed into retrieval, memory, and tool steps. The evaluation side of this literature is equally relevant. ConfAIde shows that models leak in a contextual-integrity sense---disclosing information that is appropriate in one context and not another---even when no identifier is present~\cite{mireshghallah2024confaide}, and Staab et al.\ show that LLM adversaries infer personal attributes from text that carries no explicit identifiers at all~\cite{staab2024beyond}. Both predict the gap that Section~\ref{sec:res:sem} measures across methods: removing identifier strings is not the same as removing what a reader learns.

\subsection{Privacy-Preserving Agent and Tool-Use Pipelines}
Modern LLM agents coordinate foundation models with retrieval, tools, code execution, and persistent memory~\cite{yao2022react,schick2023toolformer}. Retrieval-augmented generation and tool-use pipelines improve utility but also create additional propagation paths for sensitive content~\cite{lewis2020retrieval,mialon2023augmented}. This orchestration expands the attack and leakage surface because sensitive prompt fragments can propagate across chained tool calls and intermediate states~\cite{perez2022ignore,greshake2023not,owasp2025llm}. Existing agent frameworks are primarily optimized for capability and task completion, and typically treat privacy protection as an external concern. Our framework addresses this gap by introducing an explicit pre-inference mediation boundary that limits propagation of raw sensitive content.

\subsection{Multimodal and OCR Privacy Protection}
Multimodal systems extend privacy exposure from typed prompts to image-derived content. OCR and visual grounding can surface identifiers and sensitive attributes from screenshots, forms, receipts, and medical documents, including content users did not explicitly type~\cite{greshake2023not,carlini2023diffusion,openai2023gpt4vsystemcard}. The evaluated scope in this paper is OCR-mediated text extraction followed by prompt mediation; full VLM-loop privacy protection remains a broader related problem.

Across these strands the difference is the \emph{stage of protection}: training-time methods protect model learning, stored-text de-identification protects records after creation, prompt-security work protects the inference-time instruction channel, and this paper addresses agent-boundary mediation before sensitive content propagates into retrieval, memory, and tool arguments.

\section{Proposed Framework}
\label{sec:framework}
\subsection{System Overview}
The proposed framework acts as a privacy mediation layer between the user and a downstream LLM-based agent, with an OCR-mediated extension for image-derived text~\cite{yao2022react,schick2023toolformer}. Instead of directly sending the raw prompt $x$ to the model, the framework first analyzes the input for privacy-sensitive content, transforms the identified spans into protected surrogates, and then forwards a sanitized prompt $\tilde{x}$ to the downstream inference pipeline.

\begin{figure*}[t]
\centering
\includegraphics[width=0.95\textwidth]{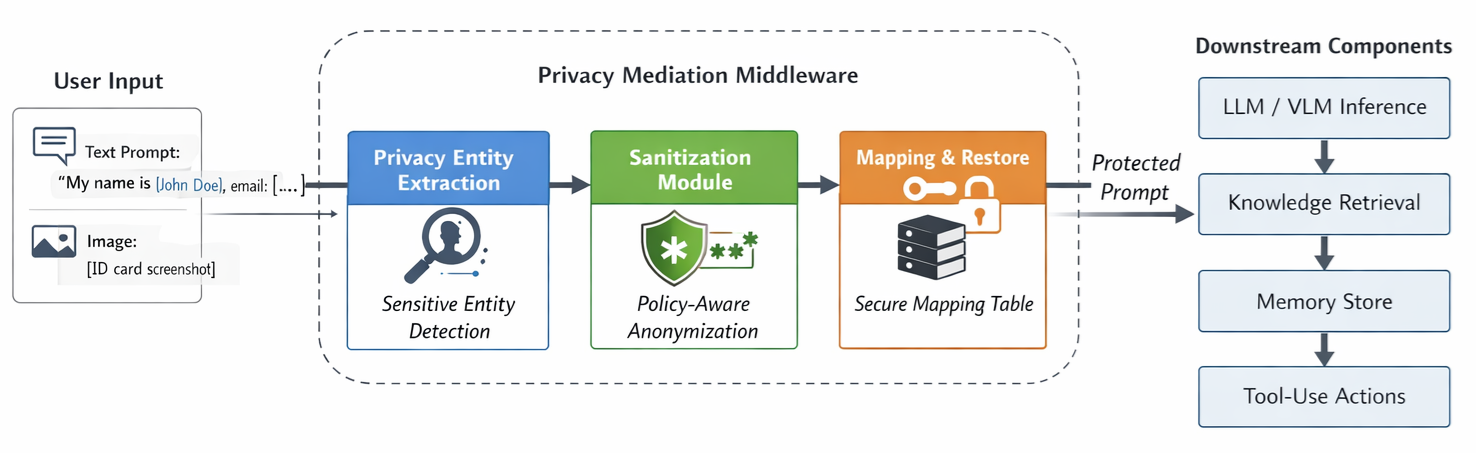}
\Description{Block diagram of the prompt-privacy mediation middleware: raw user input is intercepted, sensitive spans are detected, the configured sanitization policy is applied, and a protected prompt is forwarded to the downstream model, retrieval, memory, and tool-use components.}
\caption{Architectural overview of prompt privacy mediation for LLM agent pipelines. The middleware intercepts raw user input, detects privacy-sensitive spans, applies the configured sanitization policy, and forwards a protected prompt to the downstream model, retrieval, memory, and tool-use components. Restoration of exact values is deferred to an authorized execution boundary.}
\label{fig:workflow}
\end{figure*}

Figure~\ref{fig:workflow} shows the resulting placement: the layer sits on the single edge every later copy is derived from, so a span suppressed there is suppressed in the retrieval query, the memory write, and the tool argument that follow. The full workflow has four stages: privacy-sensitive span extraction, semantic-preserving prompt sanitization, downstream agent inference on the sanitized prompt, and optional secure restoration of selected entities under access control. This design reduces privacy exposure before model calls, retrieval, tool invocation, or logging, while preserving the semantic intent necessary for useful downstream responses.

\subsection{Privacy Entity Extraction}
Given an input prompt $x$, we first identify candidate privacy-sensitive spans:
\[
E(x)=\{(s_i,c_i,p_i)\}_{i=1}^{n},
\]
where $s_i$ denotes a detected span, $c_i$ is its privacy category, and $p_i\in[0,1]$ is a confidence score.

The detector may combine multiple signals:
\begin{itemize}
    \item rule-based recognizers for structured entities such as email addresses, phone numbers, account numbers, national identifiers, dates of birth, and postal addresses;
    \item named entity recognition for names, organizations, locations, and domain-specific entities;
    \item contextual privacy judgment for spans whose sensitivity depends on surrounding semantics;
    \item optional OCR-based extraction when privacy-bearing content appears in images or screenshots submitted as part of a multimodal prompt.
\end{itemize}
A span is marked for protection when its confidence exceeds a configurable threshold $\tau$ or when it matches a high-risk privacy policy. The output of this stage is a structured privacy annotation over the prompt.

One signal is deliberately absent from the evaluated detectors and worth naming as a gap rather than leaving to be inferred: none of them normalizes its input before matching---no Unicode confusable folding, no zero-width stripping, no surface-form canonicalization. \S\ref{sec:res:adv} measures what that costs, and the cost is the largest single number in the adversarial table. Canonicalizing before the rule and NER passes is the cheapest hardening the results here point to, and it is not what we evaluated.

This annotation serves as the control interface for downstream mode selection and surrogate construction, enabling policy-consistent behavior across heterogeneous prompt types.

\subsection{Semantic-Preserving Prompt Sanitization}
After privacy-sensitive spans are detected, the framework transforms them into protected surrogates through a policy-instantiated sanitization operator
\[
T_{\pi}(s_i,c_i)\rightarrow \hat{s}_i.
\]
The sanitized prompt is then
\[
\tilde{x}=T_{\pi}(x)=x[s_1\mapsto \hat{s}_1,\ldots,s_n\mapsto \hat{s}_n].
\]
We consider three main sanitization modes:
\begin{enumerate}
    \item \textbf{Typed placeholder replacement}, such as replacing a real person name with \texttt{<PERSON\_1>} or a phone number with \texttt{<PHONE\_1>};
    \item \textbf{Semantic abstraction}, such as replacing a full residential address with ``a residential address in a major city''---a utility-preserving mode that \S\ref{sec:res:down} shows does not control identifier exposure;
    \item \textbf{Secure symbolic mapping}, where protected spans are replaced by opaque but internally consistent tokens linked to a local mapping table under access control.
\end{enumerate}
The choice of sanitization mode depends on the privacy category, the downstream task requirement, and whether later restoration is necessary.

Once protected surrogates have been selected, the mediation layer must ensure that privacy reduction does not unnecessarily destroy downstream task semantics.

\subsection{Propagation-Aware Privacy--Utility Objective}
Prompt protection should not destroy the user's original intent, but in an agent setting it should also prevent protected surface forms from spreading across retrieval, memory, and tool boundaries before restoration is authorized. The mediation stage is therefore evaluated through the propagation-aware objective from Section~\ref{sec:problem} rather than through a single document-boundary score alone. In practice, the direct-exposure term is reported as PER, explicit boundary propagation as SPE, and utility as answer consistency, task success rate, or agent execution correctness.

Let $E(x)$ denote the set of privacy-sensitive spans in prompt $x$, and let $\widehat{E}(\tilde{x})$ denote the subset that remains directly exposed after mediation. We define the Privacy Exposure Rate (PER) as
\[
\mathrm{PER}(x,\tilde{x})=\frac{|\widehat{E}(\tilde{x})|}{|E(x)|}.
\]
For typed placeholders and secure symbolic tokens, spans count as suppressed once the original surface form is no longer directly readable in the released prompt. For semantic abstraction, the current evaluation uses a conservative direct-exposure convention: an abstracted span is still counted in $\widehat{E}(\tilde{x})$ whenever the released text retains a near-verbatim identifier cue, such as the original literal string, the same salient entity head, or another surface form that would still directly reveal the protected span without contextual inference.
For utility, let $\mathcal{S}(\cdot)$ be a task-specific normalized score (e.g., AC, TSR, or a bounded semantic quality metric). The Utility Preservation Ratio (UPR) is defined as
\[
\mathrm{UPR}(x,\tilde{x})=\frac{\mathcal{S}(f(\tilde{x}))}{\mathcal{S}(f(x))}.
\]
For prompts with $|E(x)|=0$, we define $\mathrm{PER}(x,\tilde{x})=0$ and exclude those prompts from privacy-bearing macro averages so that the metric remains well-defined at the denominator boundary. Lower PER indicates stronger direct exposure reduction, while UPR values closer to 1 indicate better retention of downstream task performance. Because PER captures residual \emph{direct} span exposure rather than inferential disclosure, it should not be interpreted as an upper bound on linkage or context-based privacy attacks.
For ordered agent stages $G=(g_1,\ldots,g_K)$ and the subset $E_k\subseteq E(x)$ that remains exposed at the boundary after stage $g_k$, we define Stage-wise Propagation Exposure (SPE) as
\[
\mathrm{SPE}_k(x,\tilde{x})=\frac{|E_k|}{|E(x)|}, \qquad g_k\in G.
\]
Here $E_k$ contains only original or near-verbatim sensitive surface forms that survive into a released boundary artifact. SPE therefore answers a narrow operational question: how much directly readable sensitive text remains visible at retrieval, memory, or tool boundaries? It does not measure whether an attacker can infer a hidden attribute from surrounding context, recover a paraphrased span, elicit a model's hidden reasoning, or compromise the restoration path.
SPE is defined per surface and is deliberately not a decay curve. The three boundaries are not a chain of successive filters applied to one another's output; they are three different projections of the same mediated prompt $\tilde{x}$. A retrieval query is a shortened, keyword-weighted rendering, a memory write is close to the full text, and a tool call is a structured extraction of the fields an API requires. Each therefore exposes a different amount of the original surface, and $\mathrm{SPE}_k$ is not in general monotone in $k$: an unprotected prompt in our reference agent exposes 48.7\% of gold spans at the retrieval boundary, 100\% at the memory boundary, and 74.8\% at the tool boundary, because the memory write copies the most text. Reading SPE as a decay would mistake the shape of the surfaces for the effect of mediation. What mediation controls is the level of every column at once, since all three are derived from $\tilde{x}$ and none of them can contain a span that $\tilde{x}$ does not.
Unlike formal DP-style guarantees, this propagation-aware mediation objective is deployment-oriented and targets practical exposure reduction before inference in agent workflows~\cite{dwork2014algorithmic,chaudhuri2011differentially,bommasani2021opportunities}.

\subsection{Sanitization Mode Selection}
Which mode a span receives should, in deployment, be a policy decision rather than a uniform setting: high-risk structured identifiers suit typed placeholders, semantically loaded spans suit abstraction, and spans whose exact value an authorized tool will later need suit symbolic mapping with restoration deferred to that boundary. Appendix~\ref{app:policy} gives a rule-level instantiation of that design space (Table~\ref{tab:pi_rules}) together with the three policy profiles the implementation ships.

We report this as design rather than as evaluated mechanism, and the distinction matters for reading the results. Every condition in this paper applies \emph{one} mode uniformly to all detected spans, so what \S\ref{sec:results} measures is modes and detectors, not a selector choosing between them per span. The one place a policy parameter was varied is the threshold sweep in Appendix~\ref{app:cppb}, and it does not move exposure at all on that surface. A selector that routes per span, and evidence that routing beats the best fixed mode, are future work rather than claims of this paper.

The mediated prompt and metadata then flow into retrieval, tools, and memory.

\subsection{Secure Mapping Table and Key Management}
Secure symbolic mapping is only meaningful if the mapping table $K$ is treated as a security-sensitive asset rather than as ordinary middleware state. In the reference design, a protected span is replaced by an opaque per-session token such as \texttt{BPSMAP-\{nonce\}-\{ctr\}}. The token carries no raw value and need not encode the entity type; type information, if needed for reasoning, is carried separately by the sanitized prompt or policy metadata. A reference deployment should keep $K$ outside model prompts, retrieval indexes, and general-purpose logs; encrypt mappings at rest using envelope encryption; issue short-lived restoration credentials scoped to one session; and record every restoration event in an append-only audit log. If available, the master wrapping key should be protected by a hardened secret manager, HSM, or TEE boundary, while expired mappings should be deleted promptly after task completion. Table~\ref{tab:mapping_surface} in Appendix~\ref{app:policy} states what each of these controls is intended to protect and where its assumptions end.

This design does not eliminate all risk. If $K$ or its restoration credentials are compromised, the confidentiality of symbolic replacements can collapse for the affected sessions. A partial table read reveals the rows that were read, while a namespace or log leak can allow cross-task linkage even without raw values. We therefore treat secure mapping as a restoration-governance mechanism that depends on explicit key-management assumptions, not as a cryptographic guarantee independent of deployment hygiene.

To move the mapping-table adversary from prose to measured behavior, Table~\ref{tab:mapcompromise} runs a capability-graded adversary against the reference per-session mechanism and measures the fraction of protected (token, value) pairs it recovers in the targeted session and across other sessions. A token-only observer and an audit-log leak recover $0\%$; a partial table read recovers exactly the rows read ($50\%$ of one session); and a leaked per-session credential recovers $100\%$ of that session only. Crucially, cross-session recovery stays at $0\%$ for every tier except master-seed compromise (T5), which is the explicit trust root. The mechanism therefore confines the blast radius of any single-session compromise to that session, and the only catastrophic failure is loss of the master secret---making the key-management assumption precise rather than rhetorical. This is a reference-design measurement under the stated isolation assumptions, not an end-to-end live-stack compromise.

\begin{table}[t]
\centering
\caption{Reference-design analysis of the secure mapping-table compromise curve, not a deployment measurement. A capability-graded adversary is run against the reference per-session mechanism; values are the measured fraction of protected (token, value) pairs recovered in the targeted session and across other sessions.}
\label{tab:mapcompromise}
\footnotesize
\setlength{\tabcolsep}{3pt}
\begin{tabularx}{\columnwidth}{>{\raggedright\arraybackslash}Xcc>{\raggedright\arraybackslash}p{1.45cm}}
\toprule
Adversary capability & Target (\%) & Cross-session (\%) & Blast radius \\
\midrule
T0 Token-only observer & 0.0 & 0.0 & Single session \\
T1 Audit-log leak & 0.0 & 0.0 & Single session \\
T2 Partial table read & 50.0 & 0.0 & Single session \\
T3 Full session table read & 100.0 & 0.0 & Single session \\
T4 Session credential leak & 100.0 & 0.0 & Single session \\
T5 Master-seed compromise & 100.0 & 100.0 & All sessions \\
\bottomrule
\end{tabularx}
\end{table}

Restoration timing is part of the security policy, not a post-processing convenience. Delaying restoration until an authorized execution boundary is what prevents raw values from re-entering retrieval, memory, and planning stages earlier in the agent pipeline.

\subsection{Agent-Level Integration}
The framework is designed for agent pipelines rather than standalone prompting. Let the downstream system be represented as $\mathcal{A}=\{f_{\mathrm{LLM}},\mathcal{Q},\mathcal{T},\mathcal{H}\}$,
where $f_{\mathrm{LLM}}$ is the foundation model, $\mathcal{Q}$ is an optional retrieval module, $\mathcal{T}$ is the toolset, and $\mathcal{H}$ is memory or logging infrastructure. Without protection, raw prompts may propagate through these components~\cite{yao2022react,schick2023toolformer,greshake2023not}. The middleware therefore intercepts user input and releases only the sanitized prompt $\tilde{x}$ together with any off-path secure mapping state needed for later restoration.

\subsection{Runtime Considerations}
Runtime is dominated by span extraction and, when enabled, OCR or contextual privacy judgment. Rule-based recognizers add little overhead, while contextual analysis and multimodal extraction add latency with model size and input length. Lightweight policies suit low-latency scenarios, whereas more expensive contextual analysis can be reserved for high-risk domains.

\subsection{OCR-Mediated Multimodal Inputs}
For multimodal settings, user input may contain both text and images. We denote such input by $x^{(m)}=(x^{(t)},x^{(v)})$. Privacy-sensitive information may appear in screenshots, scanned documents, medical reports, invoices, receipts, identity documents, whiteboards, and natural-scene text. The evaluated prototype handles this case through OCR-mediated text extraction followed by the same sanitization logic used for typed prompts. Full VLM-loop privacy validation, in which visual reasoning itself may expose or reconstruct protected content, remains outside the present evidence.

\section{Experimental Setup}
\label{sec:setup}
\subsection{Datasets and Tasks}
The evaluation covers four settings---privacy-sensitive prompt understanding, question answering over protected prompts, agent task execution under sanitized prompts, and OCR-mediated protection of image-assisted inputs---and uses four public sources with real text or real agent trajectories, and one controlled benchmark. A representative mediation example and a conceptual privacy--utility summary are given in the appendix for orientation. \emph{AI4Privacy} (PII-Masking-300K, English) provides held-out documents with gold PII spans, which are pushed through a documented deterministic reference agent. \emph{PrivacyLens}~\cite{shao2024privacylens} provides 493 agent-task trajectories with explicit sensitive items and a final tool action, on which an agent model must act. \emph{AgentDojo}~\cite{debenedetti2024agentdojo} provides sandboxed tool-executing tasks with environment-state utility checks; we use its banking and slack suites, the banking prompts carrying the exact identifiers that tools require. \emph{TAB}~\cite{pilan-etal-2022-text} is a public text-anonymization corpus used for external transfer. The controlled prompt-privacy benchmark (CPPB) is a paper-specific operating surface of 256 prompts from 32 templates across eight privacy categories, with text-only and OCR-mediated variants; it is used for policy ablations, and its construction, benchmark card, and results are reported in Appendix~\ref{app:cppb}. Public PhysioNet clinical text and the CORD, FUNSD, and SROIE document-image benchmarks support appendix-only transfer slices (Appendix~\ref{app:ocrslice}).

\paragraph{Reproducibility and code availability.}
Every record-backed table and figure here is rendered from the released records by a single table-fill pass rather than typed in, so each reported number can be regenerated from the artifact; the open-science statement lists what that artifact contains.

\subsection{Experimental Protocol}
The evaluation follows a matched-comparison protocol: every method processes the same inputs under the same downstream model and agent configuration, so that differences are attributable to mediation strategy rather than backend variance. Each source contributes a distinct measurement. AI4Privacy documents and PrivacyLens trajectories measure cross-boundary propagation of gold sensitive spans as context flows into retrieval, memory, and tool surfaces. The PrivacyLens downstream protocol measures whether the agent still completes its task and whether a marked secret reaches its final action, scored by that benchmark's own judge model and verbatim judging prompts. AgentDojo measures task utility and attack success from environment state after real tool execution, with and without prompt injection. Propagation and downstream measurements resample at the document or task level to obtain confidence intervals, and the controlled CPPB surface in Appendix~\ref{app:cppb} supplies the policy-routing and latency ablations.

\subsection{Baselines}
All methods run under the identical protocol above, so the comparison isolates mediation strategy. Besides \textbf{no protection} (raw prompts), the roster contains \textbf{regex-only} redaction of structured patterns; a \textbf{capitalization heuristic} that masks capitalized tokens, reported under its actual mechanism rather than as named-entity recognition; \textbf{spaCy NER masking} of person, organization, and location spans; \textbf{Microsoft Presidio}, installed and version-pinned, with its own recognizer set and backing model; a \textbf{Casper-style} sanitizer that merges rule and NER detections and substitutes plausible dummy values, following the design of the closest published prompt-sanitization system; an \textbf{LLM sanitizer} that prompts an 8B open-weight model to rewrite text with sensitive content removed; and the \textbf{proposed framework} in three configurations that differ only in detector---a rule-based heuristic detector, a lightweight learned span detector, and a transformer PII detector. The artifact defaults to the heuristic one so that it runs on unpacking, which is not the configuration to deploy (\S\ref{sec:res:adv}). Presidio, Casper-style, and the LLM sanitizer are the strong external comparators; the three proposed configurations isolate how much of the result comes from the propagation-control placement rather than from detector quality. Sanitization modes (typed placeholder, semantic abstraction, and symbolic mapping with early or late restoration) are evaluated on top of these detectors in the downstream experiment.

\subsection{Evaluation Metrics}
The main text reports six metrics. \emph{Privacy Exposure Rate} (PER) is residual exact-identifier exposure and \emph{Stage-wise Propagation Exposure} (SPE) is surface-form survival at each agent boundary, both defined in Section~\ref{sec:framework}. \emph{Fact Retention Rate} (FRR) is a lexical measure of how much of a sensitive fact's wording survives, and is defined below. Judged inferability is deliberately absent: it is an instrument this paper tests rather than measures with, and \S\ref{sec:res:sem} introduces it there. \emph{Restoration Success Rate} (RSR) measures authorized restoration, and \emph{Released-Context Exposure} (RCE) is the share of protected values that still appear in plaintext in the context the layer releases. RCE is a property of that released text: it is computed from what the layer emits, not from what the agent then does with it, so it does not vary with the downstream model and is not a measure of agent behavior. \emph{Attack Success Rate} (ASR) covers adaptive evasion and prompt injection. Downstream privacy and utility use the benchmarks' own scores rather than metrics of ours: PrivacyLens' judge-scored secret leakage and $0$--$3$ helpfulness, exact-identifier exposure in the agent's final action, and AgentDojo's environment-state utility and attack success. Span precision/recall/F1, the category-wise and OCR variants, the answer-consistency and task-success rubrics, and latency overhead are used only on the controlled surface and are defined in Appendix~\ref{app:cppb}.

PER is the direct-exposure metric defined in Section~\ref{sec:framework}, with $\mathrm{PER}=0$ for prompts that contain no protected spans; SPE is the corresponding surface-form survival rate at each agent boundary and is a propagation measure, not a recovery-resistance score. Two metrics sit one level above the literal span, and the distinction between them is the subject of \S\ref{sec:res:sem}. Each PrivacyLens trajectory ships a list of sensitive facts in natural language. FRR counts a fact as retained when at least half of its content words (alphabetic, length $\geq 4$, non-stopword) still occur anywhere in the released surface; it measures whether mediation removed the fact's \emph{wording}. Removing the wording is not the same as removing the fact, so we also ask the question directly: shown the released surface and one fact, with no access to the original, would a reader still learn or confidently infer it? A 27B open-weight judge from a different family than both the downstream agent and PrivacyLens' own judge answers this at temperature $0$ with a fixed seed over 200 trajectories and six methods, giving 3{,}564 judgments. A stratified 102-item subset, 17 per method, is then answered by a person seeing the same released text but neither the judge's verdict nor the method that produced the item. That subset is the reference: \S\ref{sec:res:sem} reports both automated measures against it, and neither survives.

\subsection{Implementation and Deployment Setting}
The framework is implemented as a middleware component in an agent pipeline. The privacy entity extraction stage combines deterministic rule recognizers for structured entities with NER-supported contextual extraction for unstructured spans, while the released Presidio-class comparator uses Presidio recognizers plus spaCy \texttt{en\_core\_web\_lg} for person, organization, and location detection. Placeholder design uses typed tokens (e.g., \texttt{<PERSON\_1>}, \texttt{<DATE\_1>}) to signal entity type to the downstream model, and optional secure mapping uses a locally held dictionary with randomly generated tokens as keys.

\paragraph{Models and runtime.}
Downstream agent behavior uses a 32B open-weight instruction model at temperature $0$, served locally through an OpenAI-compatible endpoint with native tool calling, and replicated on a 70B model of another family at 7k and 10k context windows for agent and judge (Appendix~\ref{app:portability}); AgentDojo tasks execute against its real sandboxed tools and are scored from resulting environment state. PrivacyLens scoring uses that benchmark's original 7B judge model with its verbatim judging prompts, so downstream leak and helpfulness numbers follow the published protocol rather than an in-house rubric. The LLM sanitizer and the semantic-abstraction mode use an 8B open-weight model from the same family. Every model is named with its quantization and digest in Appendix~\ref{app:models}; decoding parameters, library versions, and hardware are recorded per run in the released manifests. Bootstrap intervals use 1{,}000 resamples with a fixed seed. Decoding is deterministic throughout---temperature $0$, fixed seed, pinned local models---so repeating a run reproduces it: every interval here is a sampling interval over evaluation units, not across-run variance. The CPPB latency figures in Appendix~\ref{app:cppb} are serial single-request middleware overhead on a local workstation and should be read as prototype timing rather than service-scale throughput, and the CPPB repeated-run summaries use five deterministic seeds $\{17,29,43,71,101\}$ over a fixed prompt manifest.

\section{Results and Discussion}
\label{sec:results}
This section reports five measured results on real agent data: cross-boundary identifier propagation against strong baselines (\S\ref{sec:res:prop}), the surface-versus-semantic leakage gap (\S\ref{sec:res:sem}), downstream task-native privacy and utility (\S\ref{sec:res:down}), restoration timing as a security variable (\S\ref{sec:res:rest}), and behavior under adaptive evasion and prompt injection (\S\ref{sec:res:adv}). All intervals are 95\% bootstrap percentile intervals over document- or task-level resampling with 1{,}000 resamples.

\subsection{Cross-Boundary Identifier Propagation}
\label{sec:res:prop}
The central systems claim is that pre-inference mediation suppresses surface-form propagation across agent boundaries. Table~\ref{tab:measprop} measures exact gold-span survival as agent context flows into retrieval, memory, and tool surfaces on two real sources: AI4Privacy text through a documented deterministic reference agent (300 held-out documents), and real PrivacyLens agent-task trajectories~\cite{shao2024privacylens} (493 agent-privacy tasks, each shipping an executable action/observation trajectory with explicit sensitive items). Gold is the concrete surface form (user identity, verbatim e-mails, digit-run identifiers). All ten methods run under one protocol on the same inputs, including three externally defined comparators: an installed, version-pinned Microsoft Presidio; a Casper-style sanitizer combining rules, NER, and dummy-value substitution; and an LLM sanitizer prompted to rewrite the text.

\begin{figure*}[t]
\centering
\includegraphics[width=\textwidth]{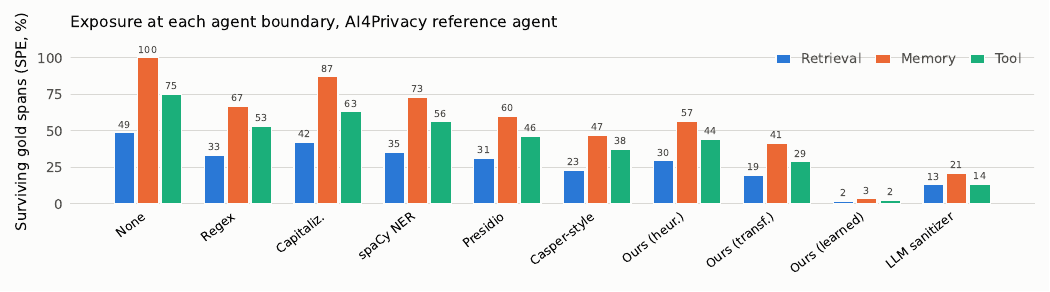}
\Description{A grouped bar chart of surviving gold spans at the retrieval, memory, and tool boundaries for ten mediation methods on AI4Privacy. The memory boundary is the highest bar for every method, and the learned detector is lowest at every boundary.}
\caption{Measured cross-boundary exposure on the AI4Privacy reference agent. Exposure is a property of each agent surface rather than a decay along a chain: the memory write copies the most text and is the highest column for every method, and mediation lowers all three columns together.}
\label{fig:propagation}
\end{figure*}

\begin{table*}[t]
\centering
\caption{Measured cross-boundary propagation under one protocol on two real sources. Mem.\ is memory-stage surface survival (SPE); PER is residual exact-identifier exposure; Ret.\ is non-sensitive text retention; FRR is the fraction of marked sensitive facts whose content words survive in the released surface, a lower bound on factual disclosure (\S\ref{sec:setup}). Brackets are 95\% bootstrap CIs; every column carries one, and Appendix~\ref{app:propints} shows them all.}
\label{tab:measprop}
\footnotesize
\begin{tabular}{lccccccc}
\toprule
& \multicolumn{3}{c}{\emph{AI4Privacy text, reference agent (300 docs)}} & \multicolumn{4}{c}{\emph{PrivacyLens agent trajectories (493 tasks)}} \\
\cmidrule(lr){2-4}\cmidrule(lr){5-8}
Method & Mem.\ SPE & PER & Ret. & Mem.\ SPE & PER & Ret. & FRR \\
\midrule
No protection & 100.0 & 100.0 & 1.00 & 100.0 & 100.0 & 1.00 & 98.8 \\
Regex-only & 66.7 & 67.4 [64.4, 70.5] & 0.98 & 12.6 & 11.8 [10.2, 13.7] & 0.99 & 98.6 \\
Capitalization heuristic & 87.1 & 87.0 [84.6, 89.4] & 0.82 & 98.9 & 99.0 [98.6, 99.3] & 0.85 & 92.4 \\
spaCy NER masking & 72.8 & 74.5 [71.0, 77.6] & 0.96 & 62.7 & 65.3 [61.5, 68.3] & 0.92 & 92.7 \\
Microsoft Presidio & 60.1 & 52.2 [48.7, 55.4] & 0.97 & 8.9 & 8.9 [7.7, 10.2] & 0.93 & 92.3 \\
Casper-style (rules + NER, dummy values) & 46.8 & 46.7 [43.4, 49.7] & 0.95 & 1.4 & \textbf{1.4 [1.0, 1.7]} & 0.91 & 91.9 \\
LLM sanitizer (8B) & 21.1 & 22.1 [19.0, 25.2] & 0.94 & 62.8 & 60.3 [56.5, 64.5] & 0.93 & 91.8 \\
Proposed (heuristic detector, artifact default) & 56.6 & 56.3 [53.1, 59.5] & 0.70 & 11.5 & 11.2 [9.6, 12.9] & 0.60 & 41.8 \\
Proposed (transformer PII detector) & 41.4 & 37.8 [34.0, 41.8] & 0.94 & 29.1 & 27.9 [24.9, 31.4] & 0.92 & 89.7 \\
Proposed (learned detector) & 3.4 & \textbf{7.4 [6.3, 8.7]} & 0.80 & 1.9 & 1.8 [1.4, 2.2] & 0.75 & 83.0 \\
\bottomrule
\end{tabular}
\end{table*}

Figure~\ref{fig:propagation} shows the boundary structure behind the table. Two results follow. First, propagation control works, and the margin over strong baselines is large where the identifiers are lexically diverse: on AI4Privacy the learned detector reaches PER 7.4\% [6.3, 8.7] against 46.7\% [43.4, 49.7] for Casper-style and 52.2\% [48.7, 55.4] for Presidio---a six- to sevenfold reduction with non-overlapping intervals---and memory-stage survival falls from 100.0\% unprotected to 3.4\%. Second, that margin is source-dependent and we do not generalize it: on PrivacyLens, where the sensitive surface is dominated by regular e-mail and name patterns, Casper-style (1.4\% [1.0, 1.7]) and the learned detector (1.8\% [1.4, 2.2]) are close, and Presidio reaches 8.9\%. The mechanism that separates the methods is not detector novelty but where mediation is applied: every method that acts before retrieval and memory writes suppresses downstream copies, while the two methods that key on capitalization or on free-form model rewriting (capitalization heuristic, LLM sanitizer) leave the majority of identifiers intact on at least one source. Retention shows the cost side: the learned detector keeps 0.80/0.75 of non-sensitive text against 0.95/0.91 for Casper-style, i.e.\ it buys its lower exposure with measurable over-redaction.

\subsection{The Surface-Form to Fact Leakage Gap}
\label{sec:res:sem}
PER and SPE ask whether a protected string survives; this subsection asks whether the \emph{fact} survives. That proves harder to measure than to ask, and the measurement problem is the first result. Three measures were applied. The lexical Fact Retention Rate counts whether a fact's content words survive (\S\ref{sec:setup}), by which every method leaving usable text retains 83.0--98.8\% of facts (Table~\ref{tab:measprop}). A judge model, shown only the released surface, was asked whether a reader would still learn the fact. On a stratified 102-item subset, 17 per method, a person answered that same question blind---without the judge's verdict, and without knowing which method produced any item.

\textbf{Both automated measures fail against the human labels.} The judge reaches Cohen's $\kappa=0.087$---chance---at 58.8\% raw agreement, calling a fact inferable in 81.4\% of items where the annotator does so in 57.8\%. The lexical measure is better and still weak, at $\kappa=0.250$. Neither can carry a claim about inferability, and we do not use them for one.

\begin{figure}[t]
\centering
\includegraphics[width=\columnwidth]{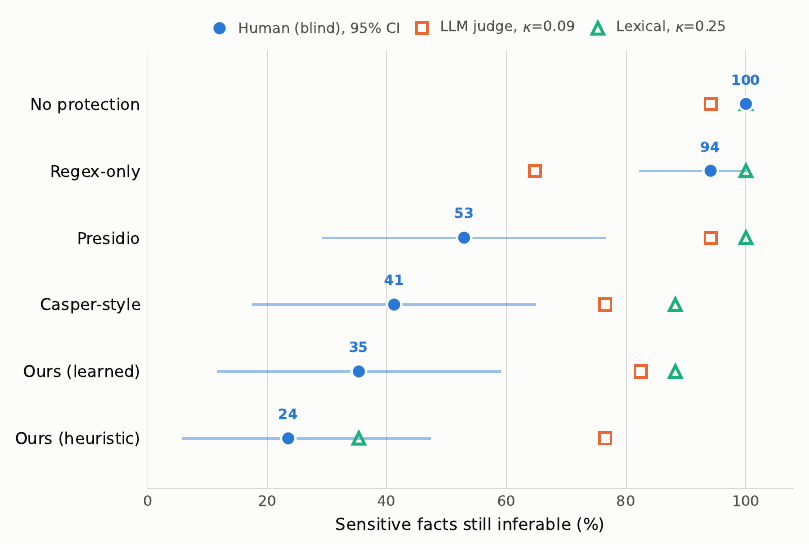}
\Description{One row per mediation method. Blue filled circles with confidence intervals show the share of sensitive facts a blind human annotator judged still inferable; these descend from 100 percent with no protection to 24 percent for the most aggressive detector. Open orange squares (the LLM judge) and open green triangles (the lexical measure) stay high and roughly flat across all methods, showing that neither proxy tracks the human labels.}
\caption{Three measures of how much a mediation method leaves inferable, on the 102-item annotated subset. The human labels (blind to method and to the judge's verdict) fall monotonically across the roster; the two automated proxies stay flat and disagree with them, at Cohen's $\kappa=0.09$ for the judge and $0.25$ for the lexical measure. Bars are 95\% bootstrap CIs on the human rate; per-method values for all three are in the released records.}
\label{fig:inferability}
\end{figure}

\textbf{The human labels reverse the finding those proxies suggested.} Figure~\ref{fig:inferability} plots all three: the proxies run near-flat across the roster while the human labels descend monotonically, from 100\% [100, 100] unprotected to 52.9\% [29.4, 76.5] for Presidio, 41.2\% [17.6, 64.7] for Casper-style, 35.3\% [11.8, 58.8] for our learned detector and 23.5\% [5.9, 47.1] for the heuristic. Each of those differs from the control by a bootstrap interval clear of zero, from $-47.1$ points [$-70.6$, $-23.5$] to $-76.5$ [$-94.1$, $-52.9$]. Mediation reduces factual inferability substantially---the opposite of what a surface metric or an unvalidated judge reports. \textbf{Two further readers, neither an author, checked those labels.} One reviewed the first 96 and moved four, leaving the ordering intact. The other rated all 102 from a blank sheet, seeing no earlier answer and no judge verdict, and agrees with the reported labels at Cohen's $\kappa=0.739$, 87.3\% raw---the first inter-rater estimate this subset supports. Against their labels the judge sits at $\kappa=0.120$ and the lexical measure at $\kappa=0.240$, so the distance between the proxies and a person is not one annotator's idiosyncrasy. Nor is it a matter of judge size: three further judges from 8B to 70B answered the same question on the same text, and the largest reaches $\kappa=0.491$---five times the reported judge, still short of two people (Appendix~\ref{app:judgecap}).

Two things follow. The layer does more than suppress identifiers: removing spans also removes much of what a reader can reconstruct, and removing more removes more. The methodological point is larger. Our judge gave a plausible, self-consistent and wrong answer, and agreed with a crude word-overlap heuristic closely enough that neither exposed the other; only human labels separated them. Concurrent work finds the same from the alignment side: LLM privacy judges agree only moderately with human annotators unless conditioned on them~\cite{tamber2026privacyalign}. Work measuring semantic leakage with a judge model---ours included, until this check---should report agreement against human annotation before treating it as a measurement.

\subsection{Downstream Task-Native Privacy and Utility}
\label{sec:res:down}
Propagation metrics measure the pipeline surface, not whether the agent still does its job or discloses the secret anyway. We therefore run the full roster through PrivacyLens' own downstream protocol on a 200-case subset: a 32B open-weight agent produces the final action, and PrivacyLens' original judge model and verbatim prompts score whether each marked secret leaked into that action and task helpfulness on its 0--3 scale. Every condition produced an executable action in 100.0\% of cases, so no method achieves privacy by refusing to act.

\begin{table*}[t]
\centering
\caption{Downstream privacy and utility on 200 PrivacyLens tasks, scored by PrivacyLens' own judge model and prompts. Leak is the fraction of tasks leaking any marked secret; Secret is the per-secret leak rate; Help.\ is the 0--3 helpfulness score; ID exp.\ is the fraction of exact identifiers reaching the final action; RCE is the share of protected values still readable in the released context, a property of that text rather than of the agent (\S\ref{sec:setup}). Brackets are 95\% bootstrap CIs.}
\label{tab:downstream}
\footnotesize
\begin{tabular}{lccccc}
\toprule
Condition & Leak (\%) & Secret (\%) & Help.\ (0--3) & ID exp.\ (\%) & RCE (\%) \\
\midrule
No protection & 38.0 [31.5, 45.0] & 19.0 [14.9, 23.4] & 2.42 [2.29, 2.55] & 13.7 [11.1, 16.4] & 0.0 \\
Typed placeholder (Presidio) & 36.5 [29.5, 43.5] & 22.7 [18.2, 27.3] & 2.25 [2.10, 2.40] & 2.8 [1.6, 4.2] & 0.0 \\
Casper-style & 28.5 [22.5, 34.5] & 14.4 [11.1, 17.8] & 1.69 [1.50, 1.86] & 2.1 [1.2, 3.2] & 0.0 \\
Typed placeholder (learned) & 29.0 [22.5, 35.5] & 15.4 [11.8, 19.3] & \textbf{2.21 [2.06, 2.35]} & 3.1 [2.0, 4.3] & 0.0 \\
Typed placeholder (transformer) & 37.5 [31.5, 44.5] & 20.3 [16.1, 25.2] & 2.10 [1.94, 2.27] & 2.7 [1.5, 3.9] & 0.0 \\
LLM sanitizer (8B) & 37.5 [30.5, 44.0] & 21.6 [16.8, 26.5] & 2.33 [2.18, 2.46] & 7.2 [5.0, 9.8] & 0.0 \\
Semantic abstraction (8B) & 34.0 [27.5, 40.5] & 18.8 [14.6, 23.0] & 2.27 [2.12, 2.40] & 12.9 [9.9, 15.9] & 0.0 \\
Symbolic mapping, late restoration & 35.0 [28.5, 41.5] & 19.3 [15.2, 23.4] & 2.26 [2.12, 2.39] & 11.6 [9.4, 14.2] & \textbf{1.6 [1.1, 2.1]} \\
Symbolic mapping, early restoration & 32.0 [25.5, 38.5] & 18.5 [14.1, 22.9] & 2.15 [1.99, 2.31] & 4.8 [3.3, 6.6] & 51.0 [48.0, 54.0] \\
\bottomrule
\end{tabular}
\end{table*}

The downstream numbers confirm the gap identified in \S\ref{sec:res:sem} under task-native scoring. Mediation is effective on the channel it targets: exact-identifier exposure in the final action falls from 13.7\% [11.1, 16.4] unprotected to 2.1--3.1\% under the three placeholder and Casper-style conditions, a four- to sixfold reduction. It is far weaker on the channel it does not target, where the shared cases call for a paired test rather than overlapping intervals: Casper-style lowers the judge-scored leak rate by 9.5 points and the learned detector by 9.0 (McNemar $p=0.009$, $0.022$), neither surviving Holm correction. That negative is agent-specific rather than general: on a 70B agent of another family, run on the same cases with the same judge, six of the eight conditions do survive Holm correction and the learned detector reaches $-18.5$ points (Appendix~\ref{app:portability}). An agent handed \texttt{<PERSON\_1>} instead of a name still discloses what that person did. That channel's instrument is the benchmark's own judge, validated against two readers in Appendix~\ref{app:pljudge}. Read across channels the evidence is uneven: exact identifiers are controlled fourfold to sixfold and hold; judge-scored leakage moves nine points and does not survive correction on the 32B agent, though six of eight conditions do on the 70B; factual inferability falls from 100\% to 24--53\% on 102 annotated items.

Utility is where the methods separate, and it reverses the ordering a privacy-only reading would suggest. Casper-style costs 0.73 helpfulness points against no protection (1.69 [1.50, 1.86] versus 2.42 [2.29, 2.55])---its dummy-value substitution produces text the agent then acts on incorrectly. Typed placeholders with the learned detector match it on leak rate (paired $p=1.0$) while retaining helpfulness 2.21 [2.06, 2.35], so on this benchmark that condition dominates Casper-style: equal privacy, materially higher utility. Semantic abstraction preserves utility well (2.27) but leaves exact-identifier exposure at 12.9\%, close to unprotected, because abstracted spans still carry the specific referent.

\subsection{Restoration Timing as a Security Variable}
\label{sec:res:rest}
The last two rows of Table~\ref{tab:downstream} isolate the one choice fully under the layer's control: detector and mapping fixed, only the restoration point moves. RCE reports what that choice controls directly---1.6\% [1.1, 2.1] of protected values stay readable in the released context under deferral against 51.0\% [48.0, 54.0] when values return during planning---and it is computed from what the layer emits, so it moves by construction and no agent behavior can contradict it. The agent-dependent measurement agrees with it. Counting gold identifiers in the action as the agent wrote it, before the authorized boundary restores anything, deferral leaves 2.7\% [1.5, 4.1] against 4.8\% [3.3, 6.6] for early restoration and 13.7\% [11.1, 16.4] unprotected: an agent that never saw a value emits it less often. Both conditions run the same 200 cases, so that comparison is paired rather than read from the marginal intervals, and paired it is resolved: $-2.1$ points [$-3.6$, $-0.8$], with a pre-restoration identifier appearing in 8.0\% of deferred final actions against 14.5\% of early-restored ones (McNemar, $1/14$ discordant, $p=0.001$). A second agent from another family reproduces it at $-2.6$ points [$-3.8$, $-1.4$]. The ID column of Table~\ref{tab:downstream} is taken after restoration, which is why deferral reads higher there (11.6\%); that column counts the disclosure the authorized boundary exists to make. Deferral costs 0.11 helpfulness points (2.26 against 2.15, overlapping) and judge-scored leakage does not separate the two, so the protection is close to free and does not extend to what the agent chooses to say.

\subsection{Adaptive Evasion and Prompt Injection}
\label{sec:res:adv}
A mediation layer must also be evaluated against an adversary who knows it is there. We report two attack surfaces. First, surface-form evasion: Table~\ref{tab:advmeas} in the appendix runs eight adaptive capabilities against the actual detectors on real AI4Privacy text, reporting attack success rate (ASR, the fraction of previously-caught spans pushed to evade). The heuristic detector the artifact defaults to is brittle across the board---zero-width insertion alone reaches 99.7\% ASR [98.9, 100.0]---so a deployable configuration must pair the framing with a learned detector. The learned detector absorbs most pure surface noise---six of the eight families stay below 1\% ASR---but two get through: case inversion at 18.4\% [13.3, 23.4] and paraphrase digit spell-out at 28.2\% [21.1, 35.6], both because they strip the shape and digit cues the detector keys on rather than merely perturbing bytes.

Measuring evasion alone understates the consequence, so we re-ran the eight attacks end-to-end on 200 documents and measured residual exposure after propagation. The learned detector stays near its clean 8.1\% PER under byte-level perturbation (8.2--13.8\%), loses ground to case inversion (19.9\%), and degrades under attacks that change token structure or spelling (45.5--72.9\%); the transformer detector fails differently, collapsing under zero-width insertion and fullwidth substitution (88.9--90.9\%). No detector in the roster is robust across all eight families, and the surviving attacks are exactly the semantic ones identified in \S\ref{sec:res:sem}.

The second attack surface is prompt injection against a live tool-executing agent. We run AgentDojo~\cite{debenedetti2024agentdojo} under its strongest published prompt-injection attack---the one that appends an instruction claiming to come from the user---across two suites, banking and slack, restricted to the injections that route user data to an attacker---the only ones a data-plane layer could affect---for 74 task/injection pairs, and compare an unprotected agent against two restoration policies (Table~\ref{tab:inject}, appendix). Mediation does not defend against this attack. Attack success is 74.3\% [64.9, 83.8] unprotected, 68.9\% [59.5, 79.7] with unrestricted restoration, and 63.5\% [52.7, 74.3] when restoration is confined to recipients named in the user's own prompt. Tested paired, confining restoration changes attack success by $-10.8$ points [$-23.0$, $+1.4$] and unrestricted restoration by $-5.4$ [$-14.9$, $+4.1$], both intervals containing zero: the ordering is consistent but not resolved at 74 pairs. What the runs do show is boundary control, on the one suite with a base for it: across banking's protected cases raw values reach the tool boundary in 0.0--8.3\% while authorized restoration succeeds 85--100\%. Only 2 slack tasks put a protected value in a tool argument, too few to rate. Injection is a control-flow attack on the instruction channel, which a data-plane layer neither should nor does prevent; what the layer changes is the \emph{value} the attacker steals, usually a placeholder rather than a live identifier.

\subsection{Deployment Interpretation and Failure Modes}
The results rank the knobs: detector choice sets the exposure floor and the over-redaction cost (\S\ref{sec:res:prop}), restoration timing decides what readers upstream of the authorized boundary see (\S\ref{sec:res:rest}), and the sanitization mode sets the utility profile. Appendix~\ref{app:deploy} reads these as a deployment ranking and states why the detector is replaceable rather than a contribution.

Four failure modes recur. Facts survive mediation more often than identifiers do (\S\ref{sec:res:sem}), and an adversary instantiated against the released surface recovers most of what the strongest detector removed: asked to fill a blanked value from four same-kind candidates, it answers 64.0\% [60.8, 67.0] correctly against a 25\% chance line on exactly the questions whose value the detector took out, and that rate barely moves across a more than eightfold range of attacker capacity (Appendix~\ref{app:attrrecovery}). The asymmetry is the sharpest statement of this paper's result: how much surface form a detector removes governs what a downstream reader can copy, and barely governs what an adversary can reconstruct; semantic evasion defeats detectors that typographic evasion does not (\S\ref{sec:res:adv}); dummy-value substitution can damage the task, costing Casper-style 0.73 helpfulness points while matching the placeholder leak rate (\S\ref{sec:res:down}); and context-dependent spans and dense OCR noise still cause category and span-boundary errors (Appendices~\ref{app:cppb} and~\ref{app:ocrslice}).

\section{Limitations}
\label{sec:limitations}
Four limitations bound this work. First, the factual-inferability labels we report come from one reader, an author, who annotated all 102 items blind to method and to the judge's verdict; two further readers, neither an author, checked them (\S\ref{sec:res:sem}). Rater reliability is no longer open---a reader working from a blank sheet agrees at $\kappa=0.739$---but at 17 items per method the rates carry wide intervals even where the ordering is clear. Second, mediation does not defend the instruction channel (\S\ref{sec:res:adv}). Third, the pipeline offers no end-to-end formal guarantee, with the scope stated in \S\ref{sec:formalscope}. Fourth, the evidence is bounded in scope---a 200-task PrivacyLens subset run on two agent models, 74 AgentDojo task/injection pairs on one, English-only text, and a template-derived benchmark (Appendix~\ref{app:cppb}). Every model evaluated is open-weight and served locally, so behavior behind a commercial endpoint and its own upstream filtering is unknown; the portability check runs between two open families instead (Appendix~\ref{app:portability}). The downstream leak rates of \S\ref{sec:res:down} are scored by PrivacyLens' own judge, which we validated rather than assumed: it agrees with two independent readers at $\kappa\approx0.51$ where they agree with each other at $0.917$ (Appendix~\ref{app:pljudge}), so we report them as a benchmark's scoring rather than as a measurement of disclosure.

\section{Conclusion}
\label{sec:conclusion}
This paper reframes prompt privacy in LLM agent pipelines as propagation control across retrieval, memory, and tool boundaries rather than one-shot document redaction, and tests it against strong baselines on real agent data. Identifier propagation is controllable: the detector sets how low exposure goes and what it costs in over-redaction, while restoration timing decides what every reader upstream of the authorized boundary sees, for a tenth of a helpfulness point. Factual disclosure falls too, from 100\% inferable unprotected to 24--53\% under mediation---but only blind human annotation shows this: a word-overlap metric and an LLM judge both reported otherwise, the judge at chance agreement. A mediation layer should therefore be deployed for what it demonstrably does---keeping literal identifiers out of retrieval, memory, tool arguments, and logs until an authorized boundary needs them---not relied on to stop an agent disclosing the facts themselves. Closing it requires mediating meaning, not surface form---the fact rather than the span, scored against the human-labelled subset released here.

\bibliographystyle{ACM-Reference-Format}
\bibliography{ref}

\appendix
\section*{Appendix}
\addcontentsline{toc}{section}{Appendix}

This appendix collects the controlled supporting tables that complement the main-text evidentiary backbone, together with the core sanitization pseudocode. It is intentionally focused on reviewer-relevant supporting evidence and reproducibility boundaries; the full set of release manifests, planning matrices, illustrative examples, and per-slice provenance cards is available in the released artifact.

\section{Notation and Reproducibility Scope}
For consistency with the main text, raw prompts are denoted by $x$, sanitized prompts by $\tilde{x}$, and detected privacy spans by $E(x)=\{(s_i,c_i)\}_{i=1}^{n}$, where $s_i$ is a candidate span, $c_i$ its privacy category, and $p_i\in[0,1]$ a confidence score. The policy-conditioned sanitization operator is $T_{\pi}$, the routing selector $\Pi_{\pi}$, the restoration policy $\rho$, and the secure mapping table $K$. Every table below is generated from the released experiment records by the same table-fill pass used for the main text. Two evidence populations are kept apart throughout: the real-data slices (AI4Privacy, PrivacyLens, AgentDojo, TAB, PhysioNet, and the public document-image benchmarks), which carry the paper's claims, and CPPB, a paper-specific controlled surface used to isolate the routing policy.

\section{Model Roster}\label{app:models}
Table~\ref{tab:models} names every model used in the evaluation with its parameter count, quantization, and the leading bytes of its digest, so that a reader can obtain the same weights rather than an approximation of them. Open-weight models were served locally through an OpenAI-compatible endpoint; nothing was routed to a hosted API. The per-run manifests in the released artifact carry the full digests alongside decoding parameters, context length, library versions, and hardware.

\begin{table}[h]
\centering
\scriptsize
\setlength{\tabcolsep}{3pt}
\caption{Models by role, with quantization and digest prefix.}
\label{tab:models}
\begin{tabularx}{\columnwidth}{@{}>{\raggedright\arraybackslash}p{2.35cm}>{\raggedright\arraybackslash}Xll@{}}
\toprule
Role & Model & Size / quant. & Digest \\
\midrule
Downstream agent & \texttt{qwen3:32b} & 32.8B / Q4\_K\_M & \texttt{cdf6a2e0} \\
Downstream agent, portability & \texttt{llama3.3:70b} & 70.6B / Q4\_K\_M & \texttt{bb2a7bfe} \\
PrivacyLens judge & \texttt{mistral:7b-\allowbreak instruct-v0.2} & 7.2B / Q4\_K\_M & \texttt{79c08d35} \\
Inferability judge & \texttt{gemma2:27b} & 27.2B / Q4\_0 & \texttt{53261bc9} \\
LLM sanitizer, abstraction & \texttt{qwen3:8b} & 8.2B / Q4\_K\_M & \texttt{a0a5ad80} \\
Transformer PII detector & \texttt{urchade/\allowbreak gliner\_multi\_\allowbreak pii-v1} & \textemdash & \textemdash \\
spaCy NER, Presidio, Casper-style & \texttt{en\_core\_\allowbreak web\_lg} & \textemdash & \textemdash \\
\bottomrule
\end{tabularx}
\end{table}

The inferability judge is deliberately from a different family than both the downstream agent and the PrivacyLens judge, so that agreement between a judge and the text it scores cannot come from shared weights. The learned span detector and the heuristic detector are trained or specified in the artifact rather than downloaded, and carry no external identifier.

\section{CPPB Construction and Data Card}
CPPB is built from 32 prompt templates spanning four prompt families---direct user requests, document-oriented instructions, retrieval-style queries, and tool-oriented agent prompts---authored from public task sources, dialogue templates, and document-understanding scenarios rather than from real user records. Each template is expanded into eight fixed variants by injecting privacy-sensitive spans at controlled positions, yielding $32\times 8 = 256$ prompts. Variants V1--V4 form the \emph{essential-privacy} subset, where the sensitive span is required to complete the task, and V5--V8 the \emph{incidental-privacy} subset, where the span is present but not task-critical; V4 and V8 are the OCR-mediated text-plus-image slice. Injected spans cover eight primary categories (person, phone, address, national identifier, financial reference, medical content, organization term, and context-dependent confidential span), balanced at 32 prompts per category and 128/128 across the two subsets. Span text, character offsets, category labels, subset membership, and modality are construction-time labels recorded in the released manifest, not model predictions; each prompt was reviewed for label correctness and for whether the injected span is genuinely essential or incidental to its stated task. Table~\ref{tab:app-examples} shows representative templates and their mediated forms across sanitization modes.

\begin{table}[t]
\centering
\caption{Representative CPPB prompt fragments and their mediated forms across privacy categories and sanitization modes, illustrating template construction and span injection.}
\label{tab:app-examples}
\scriptsize
\setlength{\tabcolsep}{3pt}
\begin{tabularx}{\columnwidth}{>{\raggedright\arraybackslash}p{1.15cm}>{\raggedright\arraybackslash}X>{\raggedright\arraybackslash}p{1.05cm}}
\toprule
Category & Raw fragment $\rightarrow$ mediated fragment & Mode \\
\midrule
Person & ``email Sarah Chen about the contract'' $\rightarrow$ ``email \texttt{<PERSON\_1>} about the contract'' & Placeholder \\
Medical & ``a patient with stage III pancreatic cancer'' $\rightarrow$ ``a patient with a serious oncological condition'' & Abstract \\
Financial & ``charge account 4012-0034-5566-7788'' $\rightarrow$ ``charge account \texttt{[TOKEN\_2fa4]}'' & Symbolic \\
Visual text & ``summarize the attached invoice image'' $\rightarrow$ ``summarize the attached invoice after protecting invoice ID and supplier address'' & Abstract \\
\bottomrule
\end{tabularx}
\end{table}

\section{Controlled CPPB Operating Surface}
\label{app:cppb}
This appendix reports the paper-specific controlled benchmark (CPPB): its construction, the matched baseline comparison, category-wise behaviour, utility, policy sensitivity, split integrity, and prototype latency. CPPB serves one purpose in this paper: it is a fully controlled surface on which the routing policy $\Pi_{\pi}$, the sanitization modes, and the category-level failure structure can be varied with known ground truth. Every claim the paper defends is measured on real agent data in the main text; the ablations below explain \emph{why} the mechanism behaves as it does there, and are read as design analysis rather than as comparative evidence.

\subsection{Benchmark Construction}
Because dedicated prompt-privacy benchmarks remain limited, this study uses a controlled prompt-privacy benchmark (CPPB). CPPB is a controlled benchmark rather than a community benchmark. We construct evaluation prompts by injecting privacy-sensitive spans into realistic user instructions drawn from public task sources, dialogue templates, and document-understanding scenarios. The inserted spans include person names, phone numbers, postal addresses, national identifiers, financial references, medical content, organization-specific terms, and image-derived text where applicable.

The prompt set contains two complementary subsets:
\begin{itemize}
    \item \textbf{Essential-privacy prompts}: prompts where privacy-bearing spans are integral to the task, requiring semantic abstraction rather than complete deletion to preserve utility.
    \item \textbf{Incidental-privacy prompts}: prompts where the sensitive span is contextual but not needed for the answer, where typed placeholder replacement is sufficient.
\end{itemize}

To reduce evaluation bias, privacy-sensitive spans are injected across multiple categories and prompt styles, including direct requests, document-oriented instructions, retrieval-style prompts, and tool-oriented agent prompts. This design helps assess whether the mediation framework generalizes across heterogeneous prompt structures rather than only a single task template. CPPB is organized along four reporting axes: prompt family, privacy category, downstream task type, and modality (text-only versus OCR-mediated text-plus-image). Table~\ref{tab:cppb_card} gives the exact benchmark accounting: 256 prompts derived from 32 templates with 8 injected variants per template, balanced across essential and incidental privacy subsets (128/128), across four prompt families and four prompt-source groups (64 each), and across eight primary privacy categories (32 each), with a 192/64 split between text-only and OCR-mediated text-plus-image prompts.

\begin{table}[t]
\centering
\caption{CPPB benchmark card and composition. This table summarizes the released benchmark composition used to probe privacy propagation across direct requests, document prompts, retrieval prompts, and tool-oriented agent workflows before downstream inference.}
\label{tab:cppb_card}
\scriptsize
\setlength{\tabcolsep}{3pt}
\begin{tabularx}{\columnwidth}{>{\raggedright\arraybackslash}p{1.65cm}X}
\toprule
Axis & Exact accounting \\
\midrule
Benchmark total & 256 prompts \\
Subsets & Essential-privacy 128 (50.0\%); Incidental-privacy 128 (50.0\%) \\
Prompt families & Direct requests 64 (25.0\%); Document-oriented 64 (25.0\%); Retrieval-style 64 (25.0\%); Tool-oriented agent 64 (25.0\%) \\
Privacy categories & Person names 32 (12.5\%); Contact details 32 (12.5\%); Postal addresses 32 (12.5\%); National/account identifiers 32 (12.5\%); Financial references 32 (12.5\%); Medical content 32 (12.5\%); Organization/project terms 32 (12.5\%); Context-dependent confidential spans 32 (12.5\%) \\
Prompt sources & Dialogue templates 64 (25.0\%); Public task sources 64 (25.0\%); Document scenarios 64 (25.0\%); Agent workflow traces 64 (25.0\%) \\
Modality & Text-only 192 (75.0\%); OCR-mediated text-plus-image 64 (25.0\%) \\
Template/variant accounting & 32 templates x 8 injected variants \\
\bottomrule
\end{tabularx}
\end{table}

These prompt families correspond to different ingress patterns into the agent boundary graph rather than to cosmetic template variation.

CPPB train/dev/test separation is deterministic and template-stratified: all eight variants of a template are co-located in one split, giving 16/8/8 templates and 128/64/64 prompts, with zero template overlap across splits. The split-integrity audit below reports what this construction does and does not support.

The released artifact includes a deterministic CPPB template inventory, prompt-level manifest, accounting summary, and train/dev/test split card, making the benchmark composition auditable as a benchmark card rather than only a count table. Even so, CPPB remains a paper-specific controlled benchmark; a broader release should still add fuller annotation instructions, sanitization policy labels, source-level provenance, and known-failure documentation, in line with datasheet-style reporting~\cite{gebru2021datasheets,liang2023helm,srivastava2023bigbench}. We therefore treat TAB~\cite{pilan-etal-2022-text}, AI4Privacy, i2b2-style clinical PHI~\cite{stubbs2015i2b2}, and CORD/FUNSD/SROIE-style OCR benchmarks~\cite{sroie2019,park2019cord} as external-transfer or supporting validation surfaces rather than as the source of the headline claim.

\subsection{Privacy Protection Effectiveness}
Under the CPPB protocol, the proposed framework reaches lower exposure than naive masking baselines while retaining higher downstream utility than generic de-identification. The baseline named ``capitalization heuristic'' here is the same one the main text reports under that name, and which the released records still label as NER-only masking: it matches capitalization, title, and location context by regular expression and runs no named-entity recognizer, so the record label overstates it. The comparison also includes an enterprise staged redaction baseline, which uses category-aware typed placeholders without semantic abstraction or boundary restoration. Structured categories such as phone numbers, e-mail addresses, dates, and identifiers are reliably detected by rule-based recognizers, while contextual categories such as organization-specific terms and medical phrases benefit from the additional contextual judgment stage~\cite{mireshghallah2023privacy,carlini2021extracting}.

\begin{table}[t]
\centering
\caption{Privacy Exposure Rate under prompt mediation methods in CPPB. Lower values indicate less residual sensitive content after mediation and therefore lower direct exposure to downstream model, memory, and tool channels before restoration. Measured on the reconstructed 256-prompt CPPB surface under the runtime named in the release manifest (qwen3:32B answering, Mistral-7B judging, both with digests and decoding parameters recorded); the original prompt set behind earlier versions of this table no longer exists.}
\label{tab:per}
\footnotesize
\begin{tabular}{lc}
\toprule
Method & PER (\%) \\
\midrule
No protection & 100.0 \\
Regex-only & 50.0 \\
Capitalization heuristic & 62.5 \\
Generic de-identification & 0.0 \\
Enterprise staged redaction & 37.5 \\
Proposed (semantic abstraction) & 27.3 \\
Proposed (utility-constrained) & 12.5 \\
\bottomrule
\end{tabular}
\end{table}

Table~\ref{tab:per} shows that naive structured redaction addresses only part of the exposure surface, because many privacy-bearing spans are contextual rather than purely pattern-based. Within this controlled surface the proposed utility-constrained setting reaches PER 12.5\% against the enterprise staged redaction baseline's 37.5\%, and the two intervals do not overlap. An earlier version of this table, measured on the original prompt set under a runtime that was never recorded, put the baseline ahead at 8.1\% against 9.3\%. Those numbers came from different prompts as well as a different backend, so we do not attribute the change to either and report only what the named runtime measures. What the surface establishes is narrow either way: CPPB scores a single mediation step at one boundary and gives no credit for the two mechanisms this paper contributes---survival across retrieval, memory, and tool boundaries, and restoration timing. It is retained because it isolates the routing policy, not because it adjudicates between systems; Section~\ref{sec:results} is where that comparison is made.

\subsection{Category-Wise Sensitive Span Analysis}
To provide finer-grained evidence, we report category-wise Sensitive Span F1 and category-wise PER for the proposed utility-constrained setting.
The category-wise breakdown orders the categories consistently: stable lexical identifiers---person names, contact fields, and structured financial or account references---are the easiest slice, medical and address spans sit in the middle, and context-dependent confidential spans are the hardest. We report the ordering rather than per-category rates, because the per-category record predates the reconstruction of \S\ref{app:cppb} and was not regenerated under the named runtime; the ordering is what the policy argument uses and it is the part the reconstruction does not disturb. This spread is not merely descriptive; it identifies the dominant residual-risk channel in deployment. In policy terms, it justifies assigning higher protection weight in~$\Pi_{\pi}$ to context-sensitive categories and confirms that aggregate PER alone can overstate maturity when category-level failure modes are concentrated.
This category-level gradient is also consistent with the structure of the extraction stage: explicit recognizers work best for stable lexical formats, whereas organization-specific and context-dependent spans remain more sensitive to ambiguity in contextual interpretation and policy calibration. Remaining deployment risk is therefore concentrated in contextual and organization-specific spans rather than in stable lexical identifiers.

\subsection{Utility Preservation}
The controlled surface uses two rubric scores. \emph{Answer Consistency} (AC) is the task-normalized agreement
\[
\mathrm{AC}=\frac{1}{N}\sum_{j=1}^{N} g(y_j,\tilde{y}_j), \qquad g(\cdot,\cdot)\in[0,1],
\]
between the outputs $y_j=f(x_j)$ and $\tilde{y}_j=f(\tilde{x}_j)$ obtained from the raw and mediated prompts, where $g$ checks task-relevant content agreement for free-form answers and exact slot or execution-state agreement for structured outputs; \emph{Task Success Rate} (TSR) is the fraction of prompts whose mediated output satisfies the template's stated task condition. Both are defined on CPPB's fixed prompt set and are used to rank mediation modes against one another under identical inputs.

An important finding is that lower PER does not automatically translate into better utility. Typed placeholder replacement maintains strong AC and TSR for tasks that do not require exact identifier values, because typed tokens preserve role-level information. Semantic abstraction is generally more stable when contextual meaning must be retained for downstream reasoning. Generic de-identification, while reducing exposure, shows lower answer consistency and task success because replacing all sensitive spans with a single generic redaction token removes task-relevant semantics.

\begin{table}[t]
\centering
\caption{Downstream utility under prompt mediation methods in CPPB. This table reports Answer Consistency and Task Success Rate; higher values indicate stronger utility preservation after pre-inference mediation. Measured on the reconstructed 256-prompt CPPB surface under the runtime named in the release manifest (qwen3:32B answering, Mistral-7B judging, both with digests and decoding parameters recorded); the original prompt set behind earlier versions of this table no longer exists.}
\label{tab:utility}
\footnotesize
\begin{tabular}{lcc}
\toprule
Method & AC & TSR \\
\midrule
No protection & 1.00 & 1.00 \\
Regex-only & 1.00 & 1.00 \\
Capitalization heuristic & 0.99 & 1.00 \\
Generic de-identification & 0.93 & 0.94 \\
Enterprise staged redaction & 0.97 & 1.00 \\
Proposed (semantic abstraction) & 0.98 & 1.00 \\
Proposed (utility-constrained) & 0.96 & 1.00 \\
\bottomrule
\end{tabular}
\end{table}

The utility results in Table~\ref{tab:utility} indicate that privacy protection should not be evaluated solely by exposure reduction. Generic de-identification removes sensitive content aggressively, but also degrades downstream answer consistency and task success: replacing every sensitive span with one generic token removes the type information a downstream model needs to keep the task well-posed. It is the only method on this surface that the utility scores separate at all. Enterprise staged redaction and the proposed utility-constrained setting are level with each other (AC 0.98 and 0.96) and both sit above generic de-identification's 0.93, which is the direction the argument needs; but the spread is small, and task success carries none of it---the judge returns success on every prompt of six of the seven methods, and only generic de-identification falls below 1.00. A saturated metric cannot rank the methods it does not separate, and we do not use it to. AC and TSR are rubric scores defined on this controlled surface; the paper's utility claims are made in Section~\ref{sec:results} with PrivacyLens' own judge model and AgentDojo's environment-state scoring.

\subsection{Policy Sensitivity Analysis for $\Pi_{\pi}$}
To characterize the sensitivity of mediation outcomes to policy configuration, we vary the detection threshold and routing strictness of~$\Pi_{\pi}$.

\begin{table}[t]
\centering
\caption{Policy sensitivity of the routing policy $\Pi_{\pi}$ in CPPB. Each profile is its own measured operating point; the threshold does not separate them on this surface. Measured on the reconstructed 256-prompt CPPB surface under the runtime named in the release manifest (qwen3:32B answering, Mistral-7B judging, both with digests and decoding parameters recorded); the original prompt set behind earlier versions of this table no longer exists.}
\label{tab:pi_sensitivity}
\footnotesize
\begin{tabular}{lccc}
\toprule
Policy profile & PER (\%) & UPR & TSR \\
\midrule
Lenient ($\tau=0.70$) & 12.5 & 0.97 & 1.00 \\
Balanced ($\tau=0.55$) & 12.5 & 0.96 & 1.00 \\
Strict ($\tau=0.40$) & 12.5 & 0.95 & 1.00 \\
\bottomrule
\end{tabular}
\end{table}

Table~\ref{tab:pi_sensitivity} reports a result we did not expect: on this surface the routing threshold does not move exposure at all. Every $\tau$ in $\{0.30,\allowbreak 0.40,\allowbreak 0.50,\allowbreak 0.55,\allowbreak 0.60,\allowbreak 0.70,\allowbreak 0.80\}$ was run as its own operating point over all 256 prompts, and all seven return PER 12.5\%; only utility moves, and only slightly (UPR 0.95 to 0.97). The cause is visible in the detector that supplies $p_i$: its confidences are bimodal, with 73\% of tokens below 0.30 and 21\% above 0.80, and none at all in $[0.70,0.80)$. Every span carrying a protected value scores above the entire range, so no setting inside it releases one. Two things follow. There is no knee to calibrate against here, and a policy panel reported at three profiles can resemble a trade-off curve when the mechanism behind it is a threshold that no prompt is near. We therefore report the flat measurement in place of the three-point reading it supersedes. The profiles in Table~\ref{tab:pi_profiles} also differ in more than $\tau$: their privacy and utility emphases are coefficient trends, which this sweep holds fixed and so does not test.

\begin{figure}[t]
\centering
\includegraphics[width=\columnwidth]{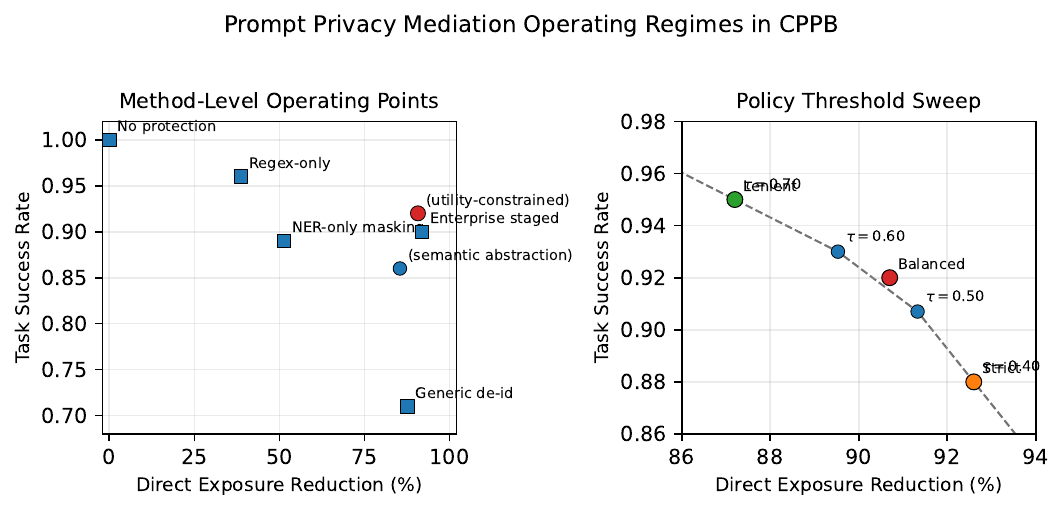}
\Description{Scatter and threshold-sweep plot of privacy-utility operating points, showing the trade-off between residual exposure and utility across policy thresholds with 95 percent confidence intervals.}
\caption{Pareto-style visual summary of privacy--utility operating regimes in CPPB, derived from Tables~\ref{tab:per}, \ref{tab:utility}, and \ref{tab:pi_sensitivity}. Left: method-level operating points using direct exposure reduction ($100-\mathrm{PER}$) and TSR. Right: a seven-point threshold sweep, every point measured over the full prompt set rather than interpolated between the named profiles.}
\label{fig:operating_points}
\end{figure}

Fig.~\ref{fig:operating_points} is a picture of operating points rather than of a frontier, and should be read as one. On this surface the proposed utility-constrained setting reaches lower exposure than the enterprise staged baseline (PER 12.5\% [8.6, 17.2] against 37.5\% [32.0, 43.8]) for a point of answer consistency (0.96 against 0.97). The right panel is flat rather than sloped: all seven thresholds return the same exposure, for the reason given above, so the sweep locates no knee to calibrate against and the figure should not be read as tracing a trade-off curve. The intervals shown are bootstrap intervals over the 256 prompts of a single deterministic pass---decoding is at temperature $0$ with a fixed seed throughout---rather than variance across repeated runs, and they do not replace an independently validated utility study.

\subsection{Repeated-Run Stability and Split-Integrity Audit}
The CPPB release has zero template overlap across its 16/8/8 train/dev/test template split, and the proposed-setting dev and test scores differ by 0.0038 PER points and 0.0001 on each of AC and TSR. That near-identity measures split construction rather than generalization: the released manifest carries prompt stubs and the bundled scores are the benchmark's original run records, so the dev/test gap says the two splits are drawn from the same 32 templates, not that the system transfers to new ones. Generalization claims in this paper are made from the real-data slices instead.

\subsection{Runtime Overhead}
Table~\ref{tab:latency} reports representative latency overhead. The policy-aware balanced mode adds 0.9\,ms mean latency (0.8\,ms at P95), and every rule-based pipeline stays under 1\,ms; the cost that dominates is the enterprise staged baseline's recognizer, at 192.2\,ms mean and 692.9\,ms at P95. Semantic abstraction is absent from the table because its mediation step is itself a model generation (4.9\,s mean), which is a property of that mode rather than middleware overhead and is not comparable with a rule-based pass. These figures are serial single-request measurements under the runtime named in the release manifest, without concurrency or a controlled prompt-length distribution, and are best read as indicative prototype costs rather than portable service-level guarantees.

\begin{table}[t]
\centering
\caption{Latency overhead of privacy mediation in CPPB. This table reports prototype runtime costs for enforcing pre-inference mediation before sensitive content can propagate into downstream agent boundaries. Measured on the reconstructed 256-prompt CPPB surface under the runtime named in the release manifest (qwen3:32B answering, Mistral-7B judging, both with digests and decoding parameters recorded); the original prompt set behind earlier versions of this table no longer exists.}
\label{tab:latency}
\footnotesize
\begin{tabular}{lcc}
\toprule
Pipeline & Mean Latency (ms) & P95 (ms) \\
\midrule
Raw prompting (no middleware) & 0.0 & 0.0 \\
Regex-only redaction & 0.9 & 0.5 \\
Capitalization heuristic & 0.1 & 0.2 \\
Generic de-identification & 0.7 & 0.7 \\
Enterprise staged redaction & 192.2 & 692.9 \\
Proposed (policy-aware balanced) & 0.9 & 0.8 \\
\bottomrule
\end{tabular}
\end{table}

The latency results indicate that practical deployment requires policy calibration. Lightweight protection is achievable with modest overhead, whereas aggressive contextual analysis improves coverage at the cost of higher latency. This trade-off is particularly important in agent settings requiring near-real-time interaction. Latency should therefore be treated as a policy-control variable and tuned jointly with acceptable residual exposure. The measured overhead is deployment-manageable for balanced profiles, but the measurements support only a prototype interpretation rather than a portable service-level claim.

\section{Threat-Aligned Supporting Evidence}
Two executed attacks against the released surface are reported here; the adaptive surface-form adversary and the secure mapping-table compromise curve are reported in the main text. With the direct surface-exposure measurement, all four adversary classes in the threat model are now met with an attack. The two here ask different questions, and the difference is the point: Table~\ref{tab:contextattack} asks the adversary to pick which case a document is, and Table~\ref{tab:attrrecovery} asks it to put back a value the release removed.

The attack is forced choice. For each of the 1{,}487 sensitive facts across all 493 PrivacyLens cases, an attacker model sees one released surface and four candidate facts---the true one and three drawn from other cases of the same sensitive-information type, so the topic alone does not answer the question. Chance is 25\%.

\begin{table}[H]
\centering
\caption{Forced-choice attack on the released surface: 1{,}487 questions per condition over 493 cases, one 32B attacker, four options, chance 25\%. Brackets are 95\% bootstrap intervals over cases. Every condition sits far above chance, which is the point of the table rather than a result about fact recovery.}
\label{tab:contextattack}
\footnotesize
\setlength{\tabcolsep}{4pt}
\begin{tabular}{lcc}
\toprule
Condition & Attack accuracy (\%) & Questions \\
\midrule
No protection & 99.3 [98.8, 99.7] & 1{,}487 \\
Typed placeholder (Presidio) & 97.6 [96.5, 98.6] & 1{,}487 \\
Typed placeholder (learned) & 95.5 [93.8, 96.9] & 1{,}487 \\
Casper-style (dummy values) & \textbf{93.3 [91.5, 94.8]} & 1{,}485 \\
\midrule
Chance & 25.0 & --- \\
\bottomrule
\end{tabular}
\end{table}

Two readings, and the second matters more. The narrow one: dummy-value substitution costs the attacker four points against a typed placeholder, 93.3\% [91.5, 94.8] against 97.6\% [96.5, 98.6] with disjoint intervals. Plausible wrong values mislead an attacker where an obvious \texttt{<PERSON\_1>} marker does not, which is a point in favour of Casper-style on this axis and against the reading that markers and dummy values are interchangeable.

The broad one is a caveat about the measurement itself. Every mediated condition stays within seven points of no protection and none approaches chance, and the reason is structural: the distractors come from other cases, so an attacker can win by recognising \emph{which} case a document is without recovering anything private from it, and mediation does not hide case identity---the tools, the institutions, the shape of the narrative all survive. This table therefore measures case identifiability, and the ordering within it is informative while its level is not. It is not evidence about whether a reader could recover the fact. That question needs a different instrument, and Table~\ref{tab:attrrecovery} is it.

\subsection{Attribute Recovery: Putting Back What the Release Removed}
\label{app:attrrecovery}
The instrument above saturates because knowing which case a document is wins the
question. This one takes that advantage away. The stem is the case's \emph{own}
sensitive fact with its most sensitive value blanked, so recognising the case
buys nothing; the adversary has to supply the missing value. The four candidates
are values of the same kind drawn from other cases---a name hidden among names,
an amount among amounts---so the type of the answer is uninformative too. Chance
is exactly 25\%, scoring is an index match, and no judge model is involved,
which matters in a paper whose own finding is that judges of this kind are
unreliable until validated. Of 1,487 marked sensitive facts, 250 carry no single
recoverable value or no same-kind pool and are dropped rather than forced into a
question with an arbitrary blank, leaving 1,237 questions over 466 cases, posed
identically in every condition.

\begin{table}[H]
\centering
\caption{Attribute recovery against the released surface: 1{,}237 questions per condition over 466 PrivacyLens cases, one 32B attacker, four same-kind candidates, chance 25\%. \emph{Recovered} is the attacker's accuracy; \emph{value present} is how often the value still appears verbatim in the released text; \emph{recovered when absent} restricts to the questions where it does not, so the answer cannot be read and must be inferred. Brackets are 95\% bootstrap intervals over cases. One reply of 4{,}948 did not parse.}
\label{tab:attrrecovery}
\scriptsize
\setlength{\tabcolsep}{2.5pt}
\begin{tabularx}{\columnwidth}{>{\raggedright\arraybackslash}Xccc}
\toprule
Condition & \makecell{Recov-\\ered (\%)} & \makecell{Value\\present (\%)} & \makecell{Recovered\\when absent (\%)} \\
\midrule
No protection & 97.7 [96.7, 98.6] & 91.7 [89.2, 93.7] & 82.5 [72.6, 90.6] \\
Typed placeholder (Presidio) & 79.0 [76.3, 81.4] & 46.4 [41.7, 50.9] & 66.6 [62.9, 70.2] \\
Casper-style (dummy values) & 75.1 [72.1, 77.9] & 44.5 [39.9, 48.9] & 59.7 [55.5, 63.6] \\
Typed placeholder (learned) & 65.3 [62.4, 68.2] & \textbf{15.2 [12.1, 18.1]} & 64.0 [60.8, 67.0] \\
\midrule
Chance & 25.0 & --- & 25.0 \\
\bottomrule
\end{tabularx}
\end{table}

The last column is the result. Mediation lowers recovery---paired on the
question, since every condition answers the same 1,237, the learned detector
costs the adversary $-32.4$ points [$-35.5$, $-29.6$], Casper-style $-22.6$
[$-25.4$, $-19.9$] and a Presidio placeholder $-18.8$ [$-21.5$, $-16.3$]---but
it does not lower it to chance, and the reason is visible in the two columns
beside it. The learned detector leaves the value verbatim in only 15.2\%
[12.1, 18.1] of releases, six times less often than a Presidio placeholder at
46.4\%. On the questions where it has removed the value, the adversary still
answers 64.0\% [60.8, 67.0] correctly against a 25\% chance line. The same
number for the Presidio placeholder is 66.6\% [62.9, 70.2] and for Casper-style
59.7\% [55.5, 63.6]. Removing three times as much of the value changes how often
it can be put back by a few points that the intervals barely separate.

We read that as the sharpest statement of this paper's central asymmetry, and
the one a deployer should carry: \emph{how much surface form a detector removes
governs what a downstream reader can copy, and barely governs what an adversary
can reconstruct}. The identifier tables measure the first and the layer does
well on them; this measures the second and no configuration we ran comes near
chance. It is the attack-side counterpart of \S\ref{sec:res:sem}, where blind
human annotation put inferability at 24--53\% under mediation rather than at
zero, and it is stronger evidence for the same conclusion because an adversary
is instantiated rather than a reader asked.

The two levels differ, and the difference is the instrument rather than a
disagreement. A forced choice hands the adversary four candidates, one of them
correct; the annotators were handed none and asked whether a reader would still
learn the fact at all. Recovering a value from a short list is easier than
recalling it unaided, so 60--67\% here and 24--53\% there bound the same
quantity from opposite sides: what a reader reconstructs with no help, and what
an adversary reconstructs when it knows what kind of thing it is looking for and
has a shortlist to check. A real adversary usually sits between the two---it
rarely has a four-item shortlist, and rarely starts from nothing either. Neither
number is the deployment quantity on its own, and we report both rather than the
one that flatters the layer.

Two limits belong with it. The value-absent column conditions on the detector's
own behavior, so its subsets differ between conditions and are not a paired
comparison; the paired numbers above are computed on all questions, where every
condition answers the same set. And the adversary is one 32B open-weight model
at temperature zero, so the level is a property of that attacker as well as of
the release. The ordering across conditions is what we rely on, and it is
consistent with the surface-form measurements: dummy-value substitution costs
the adversary $-3.9$ points [$-6.0$, $-1.6$] against a typed placeholder, the
same direction and roughly the same size as in Table~\ref{tab:contextattack},
now measured on the question the threat model actually poses.

\subsection{Does the Result Depend on How Strong the Attacker Is?}
\label{app:attrcap}
The obvious objection to a single attacker is one this paper has already
answered for judges: capacity moved judge agreement with human labels from
$\kappa=0.087$ to $0.491$ (Appendix~\ref{app:judgecap}), so a recovery rate from
one model invites the same doubt. We ran the identical attack with an 8B and a
70B attacker (Table~\ref{tab:attrcap}). Question construction is seeded, so all three were posed the same
1,237 questions---the same blanked values in the same candidate slots, verified
set-equal rather than assumed---and the runs are therefore paired on the
question rather than merely matched on protocol.

\begin{table}[H]
\centering
\caption{Recovery on the questions whose value the detector removed, as attacker capacity varies. All three attackers answer the identical questions. The 32B row is the one the main text reports. Brackets are 95\% bootstrap intervals over cases.}
\label{tab:attrcap}
\scriptsize
\setlength{\tabcolsep}{2pt}
\begin{tabularx}{\columnwidth}{>{\raggedright\arraybackslash}Xcccc}
\toprule
& \multicolumn{4}{c}{Recovered when value absent (\%)} \\
\cmidrule(lr){2-5}
Attacker & None & Presidio & Casper & Learned \\
\midrule
Qwen3 8B & 81.6 [73.1, 88.9] & 65.2 [61.4, 69.1] & 58.7 [55.1, 62.6] & 61.7 [58.5, 65.2] \\
Qwen3 32B (reported) & 82.5 [72.6, 90.6] & 66.6 [62.9, 70.2] & 59.7 [55.5, 63.6] & 64.0 [60.8, 67.0] \\
Llama 3.3 70B & 78.4 [67.9, 87.5] & 63.8 [60.0, 67.7] & \textbf{48.5 [43.8, 53.1]} & 63.8 [60.8, 66.8] \\
\midrule
Chance & 25.0 & 25.0 & 25.0 & 25.0 \\
\bottomrule
\end{tabularx}
\end{table}

Paired on the question, seven of the eight adjacent-capacity differences contain
zero. Between the 8B and the 32B every condition does, at $+1.0$ to $+2.3$
points. Between the 32B and the 70B, so do no protection ($-4.9$ [$-11.2$,
$+0.0$]), a Presidio placeholder ($-2.9$ [$-6.7$, $+0.6$]) and the learned
detector ($-0.3$ [$-3.1$, $+2.5$]). Across a more than eightfold range of attacker
capacity, how often a removed value can be put back does not measurably move.
The number the main text reports is a property of the released surface, not of
the particular attacker that read it.

The eighth difference is real and points the other way. Under Casper-style
dummy-value substitution the 70B recovers $11.5$ points \emph{fewer} than the
32B [$-15.8$, $-7.5$]. The mechanism is legible in the replies it failed to
give: 25 of its Casper answers name no candidate at all, almost all of the form
``none of these options are correct'', reasoning from the value Casper had
planted---\emph{the document actually mentions the address as Central Plaza, not
Wisteria Lane}. The stronger reader trusts a surface-plausible document more,
and a plausible wrong value exploits exactly that trust, where a
\texttt{<PERSON\_1>} marker announces itself and misleads nobody. Counting those
refusals as failures to recover, which is what they are, lowers the 70B's rate
further to 47.5\% [42.8, 51.7]; the other two attackers have at most one
unparsed reply anywhere, so no other number in this appendix moves.

For a deployer the reading is specific rather than general: against a
\emph{stronger} downstream reader, dummy-value substitution buys protection that
a typed placeholder does not, and it costs the 0.73 helpfulness points
Table~\ref{tab:downstream} already charges it. That is a trade to make
deliberately, and it runs against the intuition that a larger model is uniformly
harder to defend against.

\section{Validating the Benchmark's Own Judge}
\label{app:pljudge}
The downstream leak rates of \S\ref{sec:res:down} are scored by the judge
PrivacyLens ships, applied through its verbatim prompts. This paper's own
finding is that a judge of this kind is not evidence until it has been checked
against people, so leaving that one unchecked would have been the same mistake
with the roles swapped.

We drew 120 of the 5{,}508 recorded per-item decisions, stratified over the nine
conditions and over the judge's own verdict. The stratification matters: the
judge calls an item leaked 18.9\% of the time over the whole run, so a subset
inheriting that base rate would let it score well by answering ``no'', and raw
agreement on this subset is correspondingly not its accuracy at its natural
rate. Two readers, neither an author, answered independently from a blank sheet,
seeing the action the judge saw and the one sensitive item in question but not
its verdict. Table~\ref{tab:pljudge} reports the agreement.

\begin{table}[H]
\centering
\caption{Agreement on whether an action leaked a sensitive item, over 120 recorded decisions of PrivacyLens' own judge. Both readers worked from a blank sheet and could not see the judge's verdict or each other's answers.}
\label{tab:pljudge}
\footnotesize
\begin{tabular}{lcc}
\toprule
Comparison & Cohen's $\kappa$ & Raw agreement \\
\midrule
Judge vs.\ reader 1 & 0.500 & 75.0\% \\
Judge vs.\ reader 2 & 0.517 & 75.8\% \\
\midrule
\textbf{Reader 1 vs.\ reader 2} & \textbf{0.917} & \textbf{95.8\%} \\
\bottomrule
\end{tabular}
\end{table}

Two readings follow, and together they are more useful than either alone. The
first is that this judge is not our judge: at $\kappa\approx0.51$ it is roughly
six times the agreement our 27B inferability judge reached (\S\ref{sec:res:sem}),
so the paper's methodological claim is not that judges of this kind never work.
The second is that it is still short of a person. The readers agree with each
other at 0.917, and on the 115 items where they concur the judge dissents from
their consensus on 27---so roughly a quarter of the calls behind the reported
leak rates are not the call two readers would make. The question is not the hard
part: people answer it consistently. Consistent with that, all five items the
two readers split on come from conditions that rewrite rather than mask
(semantic abstraction, the LLM sanitizer, and one transformer-detector
placeholder); none is an unprotected or symbolic-mapping item, where both what
was released and what counts as disclosure are unambiguous.

We therefore report the downstream leak rates as this benchmark's scoring rather
than as a measurement of disclosure, and the conclusion \S\ref{sec:res:down}
draws from them---that mediation moves this channel little---is one the judge's
error rate does not overturn, since its disagreements with the readers run in
both directions rather than systematically favouring one condition.

\section{External Transfer and Learned-Baseline Comparison}
\label{app:transfer}
The public TAB text-anonymization transfer is summarized in the main text (the proposed mediator trades detection precision for lower residual exposure relative to the strongest Presidio-class baseline). Here Table~\ref{tab:lightclf} reports the AI4Privacy learned-baseline slice: a formal lightweight learned classifier is the strongest detector by recall and residual exposure, confirming that span detection on text-only PII corpora is largely solvable by simple learned means and that the contribution of this work is the propagation-control framing rather than the detector itself.

\subsection{Public Text-Only Transfer on TAB}
The appendix also reports a first executable public-benchmark transfer slice on the TAB ECHR anonymization benchmark~\cite{pilan-etal-2022-text}. Because TAB is a text anonymization corpus rather than a downstream QA or agent-task suite, this transfer slice emphasizes span precision/recall/F1, residual exposure, and non-sensitive text retention rather than CPPB-style AC/TSR.

Under the released lightweight policy-aware transfer runner, BodhiPromptShield reaches Span~F1 0.59 with PER 35.5\% and text retention 0.84. This improves on regex-only masking (0.40, 76.1\%, 1.00) and the capitalization heuristic (0.45, 61.2\%, 0.96). Compared with the stronger Presidio-class NER-fallback approximation, the transfer trade-off is mixed: BodhiPromptShield has lower precision (0.56 versus 0.76), slightly higher recall (0.63 versus 0.59), lower residual exposure (35.5\% versus 38.7\%), and lower text retention (0.84 versus 0.96). Thus the TAB slice should not be presented as a clean detection-quality win.

The precision drop is consistent with three factors. First, TAB's legal-case distribution contains names, organizations, dates, and case references whose entity boundaries do not always match the prompt-oriented categories in CPPB. Second, the transfer runner intentionally routes some quasi-identifiers and context-bearing legal references toward protection, which reduces residual exposure but counts as over-masking under TAB's annotation scheme. Third, the released lightweight detector is calibrated for prompt ingress and agent-boundary propagation rather than for maximizing span-level anonymization precision on long legal documents. We therefore interpret TAB as external evidence for a privacy--retention trade-off, not as evidence that CPPB performance directly generalizes to public legal anonymization.

The claim remains deliberately narrow. It establishes a runnable text-only external slice alongside the OCR-mediated slices below, but it does not constitute a full cross-benchmark external-baseline leaderboard.

\subsection{OCR-Mediated Transfer on Public Document Images}
\label{app:ocrslice}
The multimodal path of Section~\ref{sec:framework} is evaluated on three public document-image benchmarks: CORD receipts~\cite{park2019cord}, FUNSD forms~\cite{jaume2019funsd}, and SROIE receipts~\cite{sroie2019}. Images are read with RapidOCR (\texttt{rapidocr-\allowbreak onnxruntime} 1.2.3, default bundled multilingual detector and recognizer, default preprocessing) from pinned Hugging Face snapshots, and the recovered text is then mediated by exactly the pipeline used for typed prompts, so the slice measures whether mediation survives OCR noise rather than whether OCR itself is accurate. Table~\ref{tab:ocrslice} reports span F1 against the benchmarks' own field annotations, residual character-level exposure (PER), and non-sensitive text retention for the same baseline roster used on text.

Both text-only findings carry over. Exposure control carries over: every mediation method cuts PER far below the unprotected 100\%, and on the sparsely annotated SROIE slice the strongest baselines drive it to 1.6\%. The privacy--retention trade-off carries over and is sharper here: on CORD and FUNSD the spaCy stack reaches the lowest exposure (20.1\% and 19.9\%) by deleting two-thirds and well over half of the non-sensitive text, whereas the proposed mediation holds retention at 0.55 on both at 38.2\% and 42.7\% exposure. Detection quality, however, is not simply degraded by OCR noise: the proposed mediator reaches span F1 0.35 on CORD and 0.46 on FUNSD against 0.30 on clean AI4Privacy text and 0.59 on TAB, so the OCR slices sit inside its clean-text range rather than below it. What dominates the F1 column is annotation shape. SROIE marks 21 spans across 63 receipts, so almost every correctly protected value scores as a false positive and no method exceeds 0.02, leaving its PER column informative where its F1 column is not; FUNSD's dense field annotation instead lets Presidio and spaCy (0.56 and 0.57) outscore the proposed mediator, exactly as they do on TAB and for the same reason---this layer is not competitive as a detector, and does not claim to be. The slice is a bounded transfer result showing that mediation survives OCR-recovered text, not a claim about multimodal detection quality.

\begin{table*}[t]
\centering
\caption{OCR-mediated transfer on three public document-image benchmarks. Text is recovered with \texttt{rapidocr-onnxruntime} 1.2.3 from pinned dataset snapshots and then mediated by the same pipeline used for typed prompts. F1 is span F1 against each benchmark's field annotations, PER (\%) residual character-level exposure, and Ret.\ non-sensitive text retention. Engine, snapshot revision, preprocessing, host, and invocation are recorded per benchmark in the released runtime manifests.}
\label{tab:ocrslice}
\footnotesize
\setlength{\tabcolsep}{4pt}
\begin{tabularx}{\textwidth}{>{\raggedright\arraybackslash}Xccccccccc}
\toprule
& \multicolumn{3}{c}{CORD (100 docs, 308 spans)} & \multicolumn{3}{c}{FUNSD (50 docs, 841 spans)} & \multicolumn{3}{c}{SROIE (63 docs, 21 spans)} \\
\cmidrule(lr){2-4}\cmidrule(lr){5-7}\cmidrule(lr){8-10}
Method & F1 & PER & Ret. & F1 & PER & Ret. & F1 & PER & Ret. \\
\midrule
No protection & 0.00 & 100.0 & 1.00 & 0.00 & 100.0 & 1.00 & 0.00 & 100.0 & 1.00 \\
Regex-only & 0.07 & 95.3 & 0.63 & 0.23 & 96.4 & 0.98 & 0.02 & 68.0 & 0.89 \\
Generic de-identification & 0.11 & 88.8 & 0.62 & 0.44 & 45.9 & 0.58 & 0.02 & 9.4 & 0.46 \\
Presidio & 0.19 & 74.1 & 0.57 & 0.56 & 21.0 & 0.49 & 0.02 & 1.6 & 0.36 \\
spaCy NER & 0.35 & 20.1 & 0.34 & 0.57 & 19.9 & 0.43 & 0.02 & 1.6 & 0.32 \\
Proposed mediation & 0.35 & 38.2 & 0.55 & 0.46 & 42.7 & 0.55 & 0.02 & 9.4 & 0.46 \\
\bottomrule
\end{tabularx}
\end{table*}

\paragraph{Licensed-data-ready and synthetic i2b2 schema surface.}
For the clinical transfer path, we do not redistribute raw i2b2 notes or claim a licensed clinical slice. Instead it provides a normalization helper, a fixed PHI schema, an acquisition-tracked public i2b2-Synthea route with synthetic pilots, and a public PhysioNet-relabeled export runnable under the same wrapper without licensed archives. On that public route the release includes a seven-method comparator family over 1100 notes / 8800 PHI mentions and a held-out 100-note zero-shot pilot with three-observation stability of Span~F1 $0.55\pm0.02$, PER $29.2\pm0.7$\%, and text retention $0.81\pm0.02$. The public relabeled slice is supporting external evidence, not a substitute for an approved licensed i2b2 rerun.

\begin{table}[t]
\centering
\caption{Lightweight learned-classifier comparison on the public AI4Privacy English benchmark. A token-level class-weighted logistic-regression privacy-span detector, trained on the public training split with lexical/shape/context features, is evaluated on the held-out test split against the rule, industrial, hybrid, and proposed detectors under the same span-overlap and residual-exposure metrics.}
\label{tab:lightclf}
\scriptsize
\setlength{\tabcolsep}{3pt}
\begin{tabularx}{\columnwidth}{>{\raggedright\arraybackslash}Xccccc}
\toprule
Method & Prec. & Rec. & Span F1 & PER (\%) & Retention \\
\midrule
Regex-only & 0.76 & 0.36 & 0.49 & 63.4 & 0.98 \\
Presidio-class (+NER) & 0.23 & 0.44 & 0.30 & 52.5 & 0.70 \\
Hybrid rule+context de-id & 0.23 & 0.44 & 0.30 & 52.5 & 0.70 \\
Lightweight learned classifier & 0.36 & 0.95 & 0.53 & 7.6 & 0.79 \\
BodhiPromptShield (heuristic) & 0.23 & 0.44 & 0.30 & 52.5 & 0.70 \\
\bottomrule
\end{tabularx}
\end{table}

Cross-domain validity of the learned detector is bounded and asymmetric
(Table~\ref{tab:crossdom}): trained on generic AI4Privacy it transfers well to
clinical PhysioNet text (recall 1.00, PER 6.1\%), but a clinical-trained
detector does not generalize back to generic multi-domain PII (recall 0.38,
PER 75.4\%). The recommended generic-trained detector therefore travels across
domains, while a narrowly trained one overfits its domain---a concrete external
validity bound rather than a blanket generalization claim.

\begin{table}[t]
\centering
\caption{Measured cross-domain transfer of the lightweight learned detector between generic (AI4Privacy) and clinical (PhysioNet-relabeled) real text. Recall, Span F1, and residual exposure (PER) are measured on held-out test documents. The in-domain generic row is measured on the 300-document cross-domain sample rather than on the full 2{,}997-document AI4Privacy test split used in Table~\ref{tab:lightclf}, which is why the same detector reads Span F1 0.51 / PER 7.4\% here and 0.53 / 7.6\% there.}
\label{tab:crossdom}
\scriptsize
\setlength{\tabcolsep}{3pt}
\begin{tabularx}{\columnwidth}{>{\raggedright\arraybackslash}X>{\raggedright\arraybackslash}p{1.5cm}ccc}
\toprule
Train $\rightarrow$ Test & Setting & Rec. & F1 & PER (\%) \\
\midrule
Generic (AI4Privacy) $\rightarrow$ Generic (AI4Privacy) & in-domain & 0.96 & 0.51 & 7.4 \\
Generic (AI4Privacy) $\rightarrow$ Clinical (PhysioNet) & cross-domain & 1.00 & 0.54 & 6.1 \\
Clinical (PhysioNet) $\rightarrow$ Clinical (PhysioNet) & in-domain & 1.00 & 0.75 & 6.1 \\
Clinical (PhysioNet) $\rightarrow$ Generic (AI4Privacy) & cross-domain & 0.38 & 0.47 & 75.4 \\
\bottomrule
\end{tabularx}
\end{table}

\begin{table*}[t]
\centering
\caption{Measured adaptive adversary against the actual detectors on real AI4Privacy text (200-document test slice). ASR is the measured fraction of previously-caught sensitive spans pushed to evade detection; lower is better. Spans is the number of gold spans the attack applies to, and brackets are 95\% document-level bootstrap CIs, since spans are clustered within documents.}
\label{tab:advmeas}
\footnotesize
\setlength{\tabcolsep}{3pt}
\begin{tabularx}{\textwidth}{>{\raggedright\arraybackslash}Xrcc}
\toprule
Attack family & Spans & Heuristic ASR (\%) & Learned ASR (\%) \\
\midrule
Homoglyph substitution & 893 & 65.4 [58.0, 72.4] & 0.2 [0.0, 0.6] \\
Zero-width insertion & 893 & 99.7 [98.9, 100.0] & 0.8 [0.0, 1.8] \\
Intra-token spacing & 893 & 47.1 [31.1, 60.8] & 0.0 [0.0, 0.0] \\
Case inversion & 893 & 18.0 [2.6, 31.0] & 18.4 [13.3, 23.4] \\
Leetspeak & 893 & 33.6 [25.6, 40.1] & 0.0 [0.0, 0.0] \\
Separator injection (ids) & 472 & 0.0 [0.0, 12.1] & 0.0 [0.0, 0.0] \\
Mixed-language fullwidth & 1165 & 50.6 [43.5, 57.3] & 0.9 [0.2, 1.6] \\
Paraphrase (digit spell-out) & 472 & 65.9 [55.3, 74.8] & 28.2 [21.1, 35.6] \\
\bottomrule
\end{tabularx}
\end{table*}

\begin{table}[t]
\centering
\caption{Prompt injection on AgentDojo, pooled over the banking and slack suites (74 task/injection pairs, exfiltration-shaped injections only). ASR is attack success and Util.\ task utility; $\Delta$ is the paired change in ASR against the unprotected arm, which is the powered comparison because all arms run identical pairs. Brackets are 95\% bootstrap CIs.}
\label{tab:inject}
\footnotesize
\setlength{\tabcolsep}{3pt}
\begin{tabularx}{\columnwidth}{>{\raggedright\arraybackslash}Xccc}
\toprule
Condition & ASR (\%) & Util.\ (\%) & Paired $\Delta$ ASR (pts) \\
\midrule
No protection & 74.3 [64.9, 83.8] & 62.2 [51.4, 73.0] & --- \\
Unrestricted restoration & 68.9 [59.5, 79.7] & 56.8 [45.9, 67.6] & $-5.4$ [$-14.9$, $+4.1$] \\
User-directed restoration & 63.5 [52.7, 74.3] & 59.5 [48.6, 70.3] & $-10.8$ [$-23.0$, $+1.4$] \\
\bottomrule
\end{tabularx}
\end{table}

Widening this base is limited by suite shape rather than by compute. Banking's tasks are the ones that require an exact value at a tool boundary, so they place a protected value in a tool argument; only 2 of slack's 21 do, and the travel suite's agent never reaches a tool at all. Adding suites therefore buys task coverage rather than statistical power on restoration timing.

\section{Portability to a Second Downstream Agent}
\label{app:portability}
The downstream protocol of \S\ref{sec:res:down} was repeated end to end with a 70B open-weight agent from a different family, on the same 200 cases with the same judge model, verbatim judging prompts, decoding parameters, and seed. Table~\ref{tab:portability} reports both agents side by side; the final column gives the Holm-adjusted McNemar $p$-value for that condition against no protection on the 70B agent, over cases matched by identity rather than by marginal rate.

\begin{table*}[t]
\centering
\caption{The downstream conditions under two agents of different families, on the same 200 PrivacyLens cases. ID exp.\ is the fraction of exact identifiers reaching the final action, Leak the fraction of tasks leaking any marked secret, Help.\ the 0--3 helpfulness score. Holm $p$ is the paired McNemar test against no protection on the 70B agent, corrected across the eight conditions.}
\label{tab:portability}
\small
\begin{tabular}{@{}lccccccc@{}}
\toprule
& \multicolumn{2}{c}{ID exp.\ (\%)} & \multicolumn{2}{c}{Leak (\%)} & \multicolumn{2}{c}{Help.} & \\
\cmidrule(lr){2-3}\cmidrule(lr){4-5}\cmidrule(lr){6-7}
Condition & 32B & 70B & 32B & 70B & 32B & 70B & Holm $p$ \\
\midrule
No protection & 13.7 & 13.1 & 38.0 & 35.0 & 2.42 & 2.46 & \textemdash \\
Typed placeholder (Presidio) & 2.8 & 3.5 & 36.5 & 24.5 & 2.25 & 1.94 & 0.015 \\
Casper-style & 2.1 & 1.9 & 28.5 & 19.5 & 1.69 & 1.43 & $<$0.001 \\
Typed placeholder (learned) & 3.1 & 4.4 & 29.0 & 16.5 & 2.21 & 1.40 & $<$0.001 \\
Typed placeholder (transformer) & 2.7 & 3.5 & 37.5 & 19.5 & 2.10 & 1.87 & $<$0.001 \\
LLM sanitizer (8B) & 7.2 & 7.0 & 37.5 & 31.0 & 2.33 & 1.97 & 0.243 \\
Semantic abstraction (8B) & 12.9 & 12.0 & 34.0 & 27.0 & 2.27 & 2.11 & 0.027 \\
Symbolic mapping, late restoration & 11.6 & 9.2 & 35.0 & 25.5 & 2.26 & 2.19 & 0.022 \\
Symbolic mapping, early restoration & 4.8 & 4.7 & 32.0 & 27.5 & 2.15 & 1.90 & 0.126 \\
\bottomrule
\end{tabular}
\end{table*}

Identifier control ports. Unprotected exact-identifier exposure is 13.7\% and 13.1\%, the placeholder and Casper-style conditions land between 1.9\% and 4.4\% under both agents, and semantic abstraction stays near unprotected under both (12.9\% and 12.0\%), so the ordering the main text draws on does not depend on the agent.

Two things do not port. The first is the strength of the judge-scored leak effect: on the 32B agent only Casper-style ($-9.5$ points [2.5, 16.0], $p=0.009$) and the learned detector ($-9.0$ [2.5, 16.5], $p=0.022$) separate from no protection before correction, and neither survives it, whereas on the 70B agent six of the eight conditions survive Holm correction, the learned detector at $-18.5$ points. The negative result reported for this channel is therefore agent-specific, and the main text states it as such. The second is the utility cost of mediation, which is uniformly larger on the 70B agent: the learned detector costs $1.06$ helpfulness points there against $0.21$ on the 32B, while late symbolic restoration is the least costly configuration under both ($0.27$ and $0.16$), so the configuration the paper recommends is also the one whose cost is most stable across agents.

Released-context exposure (RCE) is deliberately absent from Table~\ref{tab:portability}. It counts protected values still visible in the released context, which is fixed by the condition and the case before the agent is called, so it cannot vary with the agent and its agreement across the two runs is arithmetic rather than replication. The restoration-timing result of \S\ref{sec:res:rest} is a mechanism-level claim and this run neither corroborates nor challenges it.

\section{Deployment Interpretation}
\label{app:deploy}
The results rank the knobs. Detector choice sets the exposure floor and the over-redaction cost (\S\ref{sec:res:prop}); restoration timing decides what readers upstream of the authorized boundary see (\S\ref{sec:res:rest}); the sanitization mode sets the utility profile, with typed placeholders giving the best measured balance. Table~\ref{tab:pi_rules} sketches how a deployment could turn that ranking into a per-span policy~\cite{nist2023airmf,eu2024aiact}; this paper does not evaluate such a policy, and the ranking above is what its results support. The invariant across profiles is that mediation must precede retrieval, memory writes, and tool-argument construction, since every downstream copy inherits whatever the first write contained.

The detector is deliberately replaceable and is not claimed as a contribution: the learned baseline is the cheapest component that makes the propagation-control layer measurable (configuration and held-out scores in \S\ref{app:transfer}), and Table~\ref{tab:measprop} shows a transformer detector and an LLM sanitizer slotting into the same protocol.

\section{Propagation Table with Intervals on Every Column}
\label{app:propints}
Table~\ref{tab:measprop} in the main text reports a bootstrap interval on PER and point estimates elsewhere, for width rather than for evidence: every column is resampled the same way, over documents for AI4Privacy and over trajectories for PrivacyLens, 1{,}000 resamples. Table~\ref{tab:propints} is the same measurement with every interval shown, rendered from the released pooled records. Nothing here is a second run; readers checking whether an ordering in the main table is resolved should read it from this one.

The retention column is the one a deployer should read first, because it prices the exposure reduction the main text leads with. Lower exposure is bought with text, and the two are not proportional. On AI4Privacy the learned detector reaches PER 7.4\% [6.3, 8.7] while keeping 0.80 [0.79, 0.81] of non-sensitive text, against Casper-style at PER 46.7\% [43.4, 49.7] and retention 0.95 [0.94, 0.95]: roughly a sixfold exposure reduction for fifteen points of text. The heuristic detector, which the artifact runs by default so that it works on unpacking, is the cautionary case, keeping 0.70 [0.68, 0.73] and 0.60 [0.58, 0.61] on the two sources while leaving more exposure than the learned detector on both---it destroys text without buying protection, which is the failure mode a retention column exists to expose. On PrivacyLens the same detector's FRR of 41.8\% [38.2, 45.5] against 83.0\% [79.7, 86.4] for the learned detector says where the text went: it removed the marked facts along with everything else. Over-redaction also has a downstream price, visible in Table~\ref{tab:downstream}: the placeholder conditions cost 0.17, 0.21 and 0.32 helpfulness points against no protection for the Presidio, learned and transformer detectors, and Casper-style, which retains the most text but substitutes plausible dummy values, costs 0.73.

\begin{table*}[t]
\centering
\caption{Cross-boundary propagation with 95\% bootstrap intervals on every column. Mem.\ is memory-stage surface survival (SPE); PER is residual exact-identifier exposure; Ret.\ is non-sensitive text retention; FRR is lexical fact retention. The AI4Privacy source carries no FRR because its documents ship no marked facts.}
\label{tab:propints}
\scriptsize
\setlength{\tabcolsep}{3pt}
\begin{tabular}{lccccccc}
\toprule
& \multicolumn{3}{c}{\emph{AI4Privacy text, reference agent (300 docs)}} & \multicolumn{4}{c}{\emph{PrivacyLens agent trajectories (493 tasks)}} \\
\cmidrule(lr){2-4}\cmidrule(lr){5-8}
Method & Mem.\ SPE & PER & Ret. & Mem.\ SPE & PER & Ret. & FRR \\
\midrule
No protection & 100.0 [100.0, 100.0] & 100.0 [100.0, 100.0] & 1.00 [1.00, 1.00] & 100.0 [100.0, 100.0] & 100.0 [100.0, 100.0] & 1.00 [1.00, 1.00] & 98.8 [98.0, 99.5] \\
Regex-only & 66.7 [63.7, 69.9] & 67.4 [64.4, 70.5] & 0.98 [0.98, 0.98] & 12.6 [10.9, 14.7] & 11.8 [10.2, 13.7] & 0.99 [0.99, 0.99] & 98.6 [97.9, 99.3] \\
Capitalization heuristic & 87.1 [84.7, 89.4] & 87.0 [84.6, 89.4] & 0.82 [0.80, 0.83] & 98.9 [98.4, 99.2] & 99.0 [98.6, 99.3] & 0.85 [0.84, 0.85] & 92.4 [90.3, 94.3] \\
spaCy NER masking & 72.8 [69.7, 75.6] & 74.5 [71.0, 77.6] & 0.96 [0.96, 0.97] & 62.7 [58.5, 66.0] & 65.3 [61.5, 68.3] & 0.92 [0.92, 0.92] & 92.7 [90.1, 95.0] \\
Microsoft Presidio & 60.1 [56.6, 63.1] & 52.2 [48.7, 55.4] & 0.97 [0.96, 0.97] & 8.9 [7.6, 10.3] & 8.9 [7.7, 10.2] & 0.93 [0.92, 0.93] & 92.3 [89.3, 94.8] \\
Casper-style (rules + NER, dummy values) & 46.8 [43.5, 49.8] & 46.7 [43.4, 49.7] & 0.95 [0.94, 0.95] & 1.4 [1.0, 1.8] & 1.4 [1.0, 1.7] & 0.91 [0.91, 0.92] & 91.9 [89.3, 94.2] \\
LLM sanitizer (8B) & 21.1 [17.6, 24.5] & 22.1 [19.0, 25.2] & 0.94 [0.92, 0.95] & 62.8 [58.5, 67.4] & 60.3 [56.5, 64.5] & 0.93 [0.92, 0.94] & 91.8 [88.6, 94.3] \\
Proposed (heuristic detector, artifact default) & 56.6 [53.1, 60.1] & 56.3 [53.1, 59.5] & 0.70 [0.68, 0.73] & 11.5 [9.7, 13.4] & 11.2 [9.6, 12.9] & 0.60 [0.58, 0.61] & 41.8 [38.2, 45.5] \\
Proposed (transformer PII detector) & 41.4 [37.6, 45.4] & 37.8 [34.0, 41.8] & 0.94 [0.93, 0.95] & 29.1 [25.8, 32.9] & 27.9 [24.9, 31.4] & 0.92 [0.91, 0.92] & 89.7 [86.5, 92.3] \\
Proposed (learned detector) & 3.4 [2.3, 4.8] & 7.4 [6.3, 8.7] & 0.80 [0.79, 0.81] & 1.9 [1.4, 2.3] & 1.8 [1.4, 2.2] & 0.75 [0.75, 0.76] & 83.0 [79.7, 86.4] \\
\bottomrule
\end{tabular}
\end{table*}

\section{Judge Capacity on the Inferability Question}
\label{app:judgecap}
The obvious reading of \S\ref{sec:res:sem} is that the judge was too small. We tested it. Three further judges spanning 8B to 70B answered the same question on the same 102 items, with the prompt, the decoding, and the released text held fixed: each judge reads the text stored in the annotation sheet, the bytes the annotators read, rather than a surface regenerated from a detector stack whose versions have since moved.

Capacity is not irrelevant, which is what the small judges alone suggest. Table~\ref{tab:judgecap} reports all four: the 70B reaches five times the reported judge's agreement. It is still only moderate where two independent readers reach substantial, and it agrees with the reported 27B judge barely at all ($\kappa=0.070$), so the two are not measuring the same thing.

The rate column says what the $\kappa$ column does not. Annotators call 57--58\% of items inferable; the 8B calls 15\%, the two mid-size judges 81--86\%, the 70B 37\%. The small judge refuses almost everything and the mid-size judges wave almost everything through. Both errors are invisible in a leak rate reported without a human check, which is the practice this paper argues against.

\begin{table}[H]
\centering
\caption{Agreement with human labels as judge capacity varies, on the 102-item subset. The 27B row is the judge the main text reports; the other three answered the identical prompt on the identical released text. Two independent readers agree at $\kappa=0.739$, shown for scale. Rightmost column: how often each rater calls a fact inferable.}
\label{tab:judgecap}
\scriptsize
\setlength{\tabcolsep}{4pt}
\begin{tabularx}{\columnwidth}{lcccc}
\toprule
Judge & Params & $\kappa$ vs.\ A1 & $\kappa$ vs.\ A3 & Calls inferable \\
\midrule
\texttt{qwen3:8b} & 8.2B & 0.223 & 0.231 & 14.7\% \\
\texttt{gemma2:27b} (reported) & 27.2B & 0.087 & 0.120 & 81.4\% \\
\texttt{qwen3:32b} & 32.8B & 0.137 & 0.173 & 86.3\% \\
\texttt{llama3.3:70b} & 70.6B & \textbf{0.491} & \textbf{0.507} & 37.3\% \\
\midrule
Annotator 1 vs.\ annotator 3 & --- & \multicolumn{2}{c}{\textbf{0.739}} & 57.8 / 56.9\% \\
\bottomrule
\end{tabularx}
\end{table}

\section{Policy Profiles and Mapping Threat Surface}
\label{app:policy}
This appendix collects the design tables referenced from Section~\ref{sec:framework}: the rule-level mode-selection design space, the policy profiles the implementation ships, and the threat surface of the secure symbolic mapping table. None of the three is a measurement. The mode-selection rules describe how a deployment \emph{could} route spans; this paper evaluates fixed modes instead, and says so where the modes are introduced.

\begin{table}[t]
\centering
\caption{Rule-level instantiation of the sanitization-mode design space. This is a design table, not a measurement: no experiment in this paper routes per span, and every reported condition applies one mode uniformly (\S\ref{sec:framework}).}
\label{tab:pi_rules}
\scriptsize
\setlength{\tabcolsep}{3pt}
\begin{tabularx}{\columnwidth}{>{\raggedright\arraybackslash}p{1.95cm}>{\raggedright\arraybackslash}p{1.45cm}>{\raggedright\arraybackslash}X}
\toprule
Routing condition & Selected mode & Rationale \\
\midrule
High-risk structured identifier; exact value not needed downstream & Placeholder & Minimizes direct exposure while preserving role/type information for reasoning. \\
Context-critical span; no exact-value tool requirement & Abstract & Retains task-relevant semantics when typed placeholders would remove too much meaning. \\
Exact value required at an authorized tool boundary & Symbolic & Keeps intermediate retrieval/memory stages opaque while allowing late restoration for execution. \\
Ambiguous contextual span under strict policy & Abstract or symbolic & Avoids under-protection when confidence is uncertain and contextual leakage risk is high. \\
\bottomrule
\end{tabularx}
\end{table}

\begin{table}[H]
\centering
\caption{The three policy profiles the implementation ships. The threshold $\tau$ is fixed numerically per profile; the privacy and utility emphases are qualitative trends encoded in the rule table above, not calibrated coefficients. No result in this paper is produced under a named profile, and the threshold sweep in \S\ref{app:cppb} does not separate them.}
\label{tab:pi_profiles}
\scriptsize
\setlength{\tabcolsep}{3pt}
\begin{tabularx}{\columnwidth}{lcccX}
\toprule
Profile & $\tau$ & Privacy emphasis & Utility/latency emphasis & Operational effect \\
\midrule
Lenient & 0.70 & Low & High & Preserves semantics more aggressively and accepts higher residual exposure when exact values are not safety-critical. \\
Balanced & 0.55 & Medium & Medium & The default mixed routing rules. \\
Strict & 0.40 & High & Low & Prioritizes suppression and delayed restoration, accepting larger utility and latency costs when risk is high. \\
\bottomrule
\end{tabularx}
\end{table}

\begin{table}[H]
\centering
\caption{Threat surface of secure symbolic mapping. The table states what the current design intends to protect and where its assumptions end.}
\label{tab:mapping_surface}
\scriptsize
\setlength{\tabcolsep}{3pt}
\begin{tabularx}{\columnwidth}{>{\raggedright\arraybackslash}p{1.55cm}>{\raggedright\arraybackslash}X>{\raggedright\arraybackslash}X}
\toprule
Surface & Intended control & Residual risk \\
\midrule
Token format & Opaque per-session token with random nonce and counter; no raw value in the token string & Tokens are not a cryptographic proof of unlinkability if session metadata leaks \\
Table visibility & $K$ is kept off prompt, retrieval, memory, and ordinary log paths & Partial table read recovers affected rows \\
Cross-task isolation & New namespace per session or task & Reuse of namespaces can link repeated entities \\
Restoration boundary & Dereference only at authorized tool or post-processing boundary & Credential compromise enables unauthorized restoration \\
Audit trail & Append-only restoration event log without raw values by default & Auditing detects misuse after the fact; it does not prevent host compromise \\
\bottomrule
\end{tabularx}
\end{table}

\section{Core Sanitization Algorithm}
Algorithm~\ref{alg:sanitize} specifies the transformation the evaluated system performs. It is written
to match the released implementation rather than the design space of \S\ref{app:policy}: the mode $m$
is fixed per condition rather than selected per span, which is why it is a parameter of the procedure
and not something the procedure computes. A deployment that routed per span would replace line~3 with
a selector over the rules in Table~\ref{tab:pi_rules}; no result in this paper does.

\begin{algorithm}[H]
\caption{Semantic-Preserving Prompt Sanitization, as evaluated}
\label{alg:sanitize}
\footnotesize
\begin{algorithmic}[1]
\Require raw prompt $x$, detected spans $E(x)=\{(s_i,c_i,p_i)\}$, confidence threshold $\tau$, sanitization mode $m\in\{\texttt{placeholder},\texttt{abstract},\texttt{symbolic}\}$, task context $\mathcal{D}$
\Ensure sanitized prompt $\tilde{x}$, secure map $K$
\Procedure{SanitizePrompt}{$x,E(x),\tau,m,\mathcal{D}$}
  \State $\tilde{x} \gets x$;\quad $K \gets \emptyset$
  \ForEach{$(s_i,c_i,p_i) \in E(x)$}
    \If{$p_i < \tau$ \textbf{and} $c_i$ is not a high-risk category}
      \State \textbf{continue} \Comment{span released unprotected}
    \EndIf
    \If{$m = \texttt{placeholder}$}
      \State $\hat{s}_i \gets \textproc{TypedPlaceholder}(c_i)$
    \ElsIf{$m = \texttt{abstract}$}
      \State $\hat{s}_i \gets \textproc{SemanticAbstract}(s_i,c_i,\mathcal{D})$
    \Else
      \State $t_i \gets \textproc{SecureToken}(s_i,c_i)$;\quad $K[t_i] \gets s_i$
      \State $\hat{s}_i \gets t_i$
    \EndIf
    \State $\tilde{x} \gets \tilde{x}[s_i \mapsto \hat{s}_i]$
  \EndFor
  \If{$\textproc{ContainsVisualInput}(x)$}
    \State $\tilde{x} \gets \textproc{AttachSanitizedVisualChannel}(\tilde{x},E(x),m)$
  \EndIf
  \State \Return $(\tilde{x}, K)$
\EndProcedure
\end{algorithmic}
\end{algorithm}

Restoration is the inverse and is deliberately not part of this procedure: $K$ is dereferenced only at
an authorized boundary, and \emph{when} that happens is the variable \S\ref{sec:res:rest} measures.

\end{document}